\documentclass[11pt,a4paper]{article}

\usepackage[english]{babel}
\usepackage[utf8]{inputenc}
\usepackage{amsfonts}
\usepackage{amsmath}
\usepackage{amssymb}
\usepackage{graphicx}
\usepackage{epsfig,epsf}
\usepackage{bm}
\usepackage{comment}
\usepackage{verbatim} 
\usepackage{booktabs}
\usepackage{array,multirow}
\usepackage{slashed}
\usepackage{xcolor}
\usepackage[font=small]{caption}
\usepackage{float}
\usepackage{blindtext}
\usepackage{placeins}
\usepackage{bbm}
\usepackage{arydshln} 
\usepackage{cite}
\usepackage{soul}
\usepackage{xfrac}
\usepackage{slashed}
\allowdisplaybreaks

\usepackage[numbers, sort&compress]{natbib}
\usepackage[numbers, sort&compress]{natbib}
\newlength{\bibitemsep}
\newlength{\bibparskip}
\let\oldthebibliography\thebibliography
\renewcommand\thebibliography[1]{%
  \oldthebibliography{#1}%
  \setlength{\parskip}{\bibitemsep}%
  \setlength{\itemsep}{\bibparskip}%
}

\def\linkcolor{cyan!70!black}
\usepackage[
colorlinks=true
,urlcolor=\linkcolor
,anchorcolor=\linkcolor
,citecolor=\linkcolor
,filecolor=\linkcolor
,linkcolor=\linkcolor
,menucolor=\linkcolor
,linktocpage=true
,pdfproducer=medialab
,pdfa=true
]{hyperref}

\usepackage{geometry}
\newenvironment{Eqnarray}{\arraycolsep 0.14em\begin{eqnarray}}{\end{eqnarray}}
\newcommand{\ba}{\begin{Eqnarray}}
\newcommand{\ea}{\end{Eqnarray}}
\newcommand{\be}{\begin{equation}}
\newcommand{\ee}{\end{equation}}
\newcommand{\bal}{\begin{aligned}}
\newcommand{\eal}{\end{aligned}}
\newcommand{\bea}{\begin{eqnarray}}
\newcommand{\eea}{\end{eqnarray}}
\newcommand{\ben}{\begin{enumerate}}
\newcommand{\een}{\end{enumerate}}
\newcommand{\bit}{\begin{itemize}}
\newcommand{\eit}{\end{itemize}}
\newcommand{\bde}{\begin{widetext}}
\newcommand{\ede}{\end{widetext}}

\newcommand{\gmuu}{\gamma^\mu}
\newcommand{\gnuu}{\gamma^\nu}

\newcommand{\gnud}{\gamma_\nu}

\renewcommand{\(}{\left(}
\renewcommand{\)}{\right)}
\renewcommand{\[}{\left[}

\newcommand{\abs}[1]{\left| #1 \right| }
\newcommand{\mathsym}[1]{{}}

\definecolor{ao}{rgb}{0.0, 0.5, 0.0}
\definecolor{bostonuniversityred}{rgb}{0.8, 0.0, 0.0}

\newcommand{\minitab}[2][l]{\begin{tabular}{#1}#2\end{tabular}}

\DeclareUnicodeCharacter{2212}{-}
\pdfoutput=1

\begin{document}
\begin{titlepage}

\vspace*{-1.0truecm}
\begin{flushright}
IFT-UAM/CSIC-22-98
 \end{flushright}
\vspace{0.8truecm}

\begin{center}
\vspace*{1cm}

{\LARGE\bf Collider Searches for Heavy Neutral Leptons: \\[1ex]
beyond simplified scenarios
}

\vspace*{0.8cm}

{\bf  Asmaa~Abada$^{a\!}$\footnote[1]{asmaa.abada@ijclab.in2p3.fr}, Pablo Escribano$^{b}$\footnote[2]{pablo.escribano@ific.uv.es}, Xabier Marcano$^{c}$\footnote[3]{xabier.marcano@uam.es} and Gioacchino Piazza$^{a}$\footnote[4]{gioacchino.piazza@ijclab.in2p3.fr}}

\vspace*{.5cm}
$^{(a)}$P\^ole Th\'eorie, Laboratoire de Physique des 2 Infinis Ir\`ene Joliot Curie (UMR 9012)\\
CNRS/IN2P3,
15 Rue Georges Clemenceau, 91400 Orsay, France

\vspace*{.5cm} 
$^{(b)}$
Instituto de F\'{\i}sica Corpuscular (CSIC-Universitat de Val\`{e}ncia), \\
 Catedr\'atico Jos\'e Beltr\'an 2, E-46980 Paterna (Valencia), Spain

\vspace*{.5cm} 
$^{{(c)}}$ Departamento de F\'{\i}sica Te\'orica and Instituto de F\'{\i}sica Te\'orica UAM/CSIC,\\
Universidad Aut\'onoma de Madrid, Cantoblanco, 28049 Madrid, Spain

\vspace*{1cm}
\begin{abstract}
\noindent
With very few exceptions,
the large amount of available experimental bounds on heavy neutral leptons - HNL - have been derived relying on the assumption of the existence of a single (usually Majorana) sterile fermion state that mixes with only one lepton flavour. 
However, most of the  extensions of the Standard Model involving sterile fermions predict the existence of several HNLs, with complex mixing patterns to all flavours. 
Consequently, most of the experimental bounds for HNLs need to be recast before being applied to a generic scenario. 
In this work, we focus on LHC searches of heavy neutral leptons and discuss how to reinterpret the available bounds when it comes to consider mixings to all active flavours, not only in the case with a single HNL, but also in the  case when  more heavy neutral leptons are involved. 
In the latter case, we also consider the possibility of interference effects and show how the bounds on the parameter space should be recast.

\end{abstract}
\end{center}

\end{titlepage}

\tableofcontents

\section{Introduction}

  Generating neutrino masses and their mixing as observed in neutrino oscillation phenomena requires to go for beyond the Standard Model of Particles (BSM). Many options are presently explored as extensions of the Higgs and/or gauge sectors, most of the time with new fields within the particle content. 
 In particular heavy neutral fermions, such as right-handed neutrinos $\nu_R$, are often present as building blocks of several neutrino mass generation mechanisms. 
 For instance, at least two $\nu_R$ are required to accommodate light neutrino masses via the type-I seesaw mechanism~\cite{Minkowski:1977sc,Yanagida:1979as,Glashow:1979nm,Gell-Mann:1979vob,Mohapatra:1979ia}.
 Moreover, in several variants of the type-I seesaw realized at low scale, other sterile (from SM gauge interactions) fermions $\nu_S$ are considered, as in the case for the Inverse~\cite{Schechter:1980gr, Gronau:1984ct, Mohapatra:1986bd} and Linear~\cite{Barr:2003nn, Malinsky:2005bi} seesaw mechanisms;  
these variants allow to have large neutrino Yukawa couplings with a comparatively low seesaw scale, potentially within reach at colliders. 
 From now on, we will refer to these states ($\nu_{R,S}$) as Heavy Neutral Leptons (HNL).

Due to the presence of HNLs, the charged and neutral currents are  modified, with the leptonic mixing matrix  encoding now not only the PMNS mixing matrix~\cite{Pontecorvo:1957cp,Maki:1962mu}, but also the active-HNL mixings $U_{\alpha N}$, $\alpha=e,\mu,\tau$. 
With these modifications and depending on the mass scale of these new neutral leptons, one expects an impact on numerous observables and thus to abundant constraints
on the plane ($M_N, |U_{\alpha N}|^2$) (see Refs.~\cite{Atre:2009rg,Abada:2017jjx,Bolton:2019pcu} and references therein).

In this work we will be interested in HNL searches at high-energy colliders, mainly at the LHC due to its current extensive program dedicated to HNL searches, see for instance~\cite{Agrawal:2021dbo}, and the numerous dedicated works and analyses~\cite{Pilaftsis:1991ug,delAguila:2007qnc,delAguila:2008cj,delAguila:2008hw,Nemevsek:2011hz,BhupalDev:2012zg,Cely:2012bz,Helo:2013esa,Dev:2013wba,Das:2014jxa,Alva:2014gxa,Izaguirre:2015pga,Deppisch:2015qwa,Maiezza:2015lza,Banerjee:2015gca,Arganda:2015ija,Das:2015toa,Gago:2015vma,Accomando:2016rpc,Degrande:2016aje,Mitra:2016kov,Ruiz:2017yyf,Dube:2017jgo,Caputo:2017pit,Antusch:2017hhu,Das:2017pvt,Das:2017gke,Deppisch:2018eth,Liu:2018wte,Cottin:2018kmq,Cottin:2018nms,Dib:2018iyr,Nemevsek:2018bbt,Abada:2018sfh,Pascoli:2018rsg,Antusch:2018bgr, Pascoli:2018heg,Drewes:2019fou,Liu:2019ayx,Jones-Perez:2019plk,Fuks:2020att,Tastet:2021vwp}.
With very few exceptions, the large amount of available HNL bounds have been derived relying on the assumption of the existence of a single (usually Majorana) HNL that mixes with only one lepton flavour. 
However, most of the BSM scenarios involving new neutral leptons address the lepton mixing as a whole, as it impacts flavour physics studies 
and lepton properties - Dirac or Majorana nature for neutrinos -  with different lepton number conserving/violating processes. 
The mixing pattern in these scenarios is expected to be quite complex, so applying the bounds from negative searches on the HNL space of parameters derived in the context of simplified hypotheses (with only one HNL which mixes to only one active flavour) does not seem adequate.
Indeed, as we will see, using these limits directly will in general overconstrain the parameter space. 
Consequently, most of the experimental bounds for HNL need to be recast before being applied to a generic BSM scenario.

 The motivation for reinterpreting LHC bounds in general is a well-established topic, see for instance Ref.~\cite{LHCReinterpretationForum:2020xtr} and references therein.
 In the context of HNL searches, the reinterpretation of the obtained  bounds on the HNL mixings to active flavours has been  addressed  in previous works, as in for instance Ref.~\cite{Abada:2018sfh} discussing searches for Heavy Neutral Leptons with Displaced Vertices, and more recently in Ref.~\cite{Tastet:2021vwp} focusing on HNLs decaying promptly to a tri-lepton final state. 
In the latter,  the single-flavour mixing results obtained by ATLAS have been recast to a low-scale seesaw model with a pseudo-degenerate pair of HNLs, the most minimal and simple  extension in order to accommodate neutrino oscillation data (the lightest neutrino being massless). 
Due to the simplicity of this model, the active neutrino masses and mixings determine the flavour pattern of the HNLs~\cite{Ibarra:2003up,Gavela:2009cd}, and it is possible to define benchmark points beyond the single-flavour scenario~\cite{Drewes:2022akb}.
While being a very interesting scenario, as it is probing the parameter space connected to light neutrino masses, this approach has the drawback of being model dependent.
For example, considering  other sources for  light neutrino masses\footnote{This would actually be needed in order to relax the hypothesis of the lightest active neutrino being massless and thus have in general 3 massive active neutrinos.}, such as additional HNLs not necessarily within the LHC range, could spoil the correlation between light and heavy sectors that motivated the definition of these scenarios. 

For this reason, in this work we will follow a different approach. 
We will instead work with physical HNL states with independent mixings and masses, with the motivation of covering every scenario that could be realized at generic BSM models. 
We will also discuss how to go beyond the simplest single-flavour mixing scenario, however we will not attempt to explain light neutrino masses and mixings. 
This idea is actually the most straightforward extension to what is usually assumed at LHC searches. 
In doing so, we will discuss what would be the most relevant quantities to be bounded experimentally in order to easily reinterpret the results.

Furthermore, we will also extend the study to the case where more than one HNL is present and, in particular, taking into account possible interference effect when at least two heavy neutral leptons are in the same mass regime (nearly degenerate or possibly forming a pseudo-Dirac neutrino pair).  
This possible interference effect can lead to different bounds on the active-sterile mixings, see for instance~\cite{Das:2017hmg,Abada:2019bac,Najafi:2020dkp}.
 
 This work is organised as follows: after having thoroughly discussed the status of high-energy collider searches of HNL in Section \ref{sec:status}, a first insight of going beyond the single active-sterile mixing approximation is examined in Section~\ref{sec:flavour}.
 In this section, we also discuss how one would reinterpret experimental data from branching ratios to limits on the model parameter space, {\it i.e.}~HNL masses and mixings.  In Section~\ref{sec:2HNL}, we consider the generic case of having more than one HNL and discuss possible changes in the interpretation of the bounds on HNL, in particular when the interference effects can be relevant. 
Final comments  are collected in Section~\ref{sec:concs}.
Further details for deriving the relevant amplitudes are collected in the Appendices.

\section{Status of HNL searches at high-energy colliders}\label{sec:status}

Heavy Neutral Leptons can be searched for in a wide variety of processes and experiments,  the HNL mass being the key ingredient to decide which is the optimal one. 
HNLs lighter than the GeV scale can lead to signatures in nuclear $\beta$ decays or in  leptonic or semileptonic meson and tau decays, while  heavy HNL above the TeV are better explored indirectly by electroweak (EW) precision observables or rare flavour processes. 
For a detailed review of all these signals and experimental status, see for instance Refs.~\cite{Atre:2009rg,Abada:2017jjx,Bolton:2019pcu}.

Here, we are interested in the intermediate regime, with HNL masses $M_N$ ranging from  few to hundreds of GeVs.
Such HNLs could be directly produced at high-energy colliders and their lifetimes are usually short enough to decay within the detectors, so we could discover them looking for their decay products. 
As these are weak processes, the Standard Model boson masses obviously define the relevant scale to be compared to. 
Along this work, we will refer as {\it light\,\footnote{Light in the context of high-energy collider searches.} HNL} to those with masses lighter than the $W$ boson, and as {\it heavy HNL} to the ones with $M_N>M_W$.
Extensive reviews about HNL searches at colliders can be found for instance in Refs.~\cite{Cai:2017mow,Batell:2022ubw,Abdullahi:2022jlv}. 
Here we just summarize and update the list of experimental analyses, introducing at the same time the most relevant aspects that we will use in our discussion in the next sections.

As in any collider search looking for heavy unstable particles, we need to  consider both the production and the decay channels of the HNLs. At a hadronic collider such as the LHC, the main production channel comes from Drell-Yan $W$ and $Z$ bosons, 
\begin{equation}
pp \to W^{(*)} \to N\ell^\pm
\quad {\rm and} \quad
pp \to Z^{(*)} \to N\nu\,,
\end{equation}
where the gauge bosons could be on- or off-shell, depending on whether  the HNL is lighter or heavier than the $W$ or $Z$  bosons.
Additional production channels could also arise from the Higgs ($H$) boson decays, which could be motivated in several models providing large neutrino Yukawa couplings.
Unfortunately, Higgs bosons are produced less abundantly than weak bosons, so they are usually neglected.
Moreover the $W$ channel has the additional prompt charged lepton that can help triggering the process and reducing backgrounds, and thus experimental searches focused mostly on this channel. 
Nevertheless, it is worth mentioning that for very heavy masses, at around the TeV scale, vector boson fusion  channels such as $W\gamma$ or $WW$ become important and could even dominate the production of HNLs~\cite{Dev:2013wba,Alva:2014gxa,Fuks:2020att}.
Indeed, the latest CMS analysis~\cite{CMS:2018jxx} already included the $W\gamma$ channel in order to enhance their sensitivity to high HNL masses.

After being produced, a HNL of several GeVs, but still lighter than $M_W$, will decay dominantly via off-shell $W$ or $Z$ bosons to a 3-body final state
\begin{align}
N&\to \ell^\pm_\alpha jj\,,\\
N&\to \ell^\pm_\alpha \ell^\mp_\beta \nu_\beta\,,\\
N&\to\nu_\alpha jj\,,\\
N&\to\nu_\alpha\ell_\beta^\pm\ell_\beta^\mp\,,\\
N&\to 3\nu\,.
\end{align}
On the other hand, if $M_N$ is above the EW scale, the dominant decays will be to on-shell $W, Z$ and $H$ bosons, {\it i.e.}~$N\to\ell^\pm W^\mp, \nu Z, \nu H$.
These 2-body decays will be followed by the decay of the heavy bosons, leading at the end to the same final states as before.
Nevertheless, it is important to keep in mind that the kinematics in these two mass regimes will be different.

\begin{figure}[t!]
\begin{center}
\includegraphics[width=.5\textwidth]{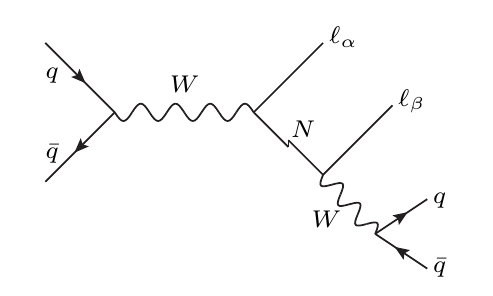}
\caption{Drell-Yann HNL production leading to a dilepton signature. 
The thunder-shaped arrow indicates that the HNL could be of Dirac or of Majorana nature, short- or long-lived. }\label{DiagHNLatLHC}
\end{center}
\end{figure}

Combining both production and decay channels, we get a full process such as the example shown in Fig.~\ref{DiagHNLatLHC}.
Depending on the relative size of $M_N$ and $M_W$, either the first or the second $W$ boson will be on-/off-shell, distinguishing the light and heavy HNL regimes.
A full catalogue of HNL signatures, combining the different production and decay processes, can be found for instance in Ref.~\cite{Antusch:2016ejd}.
Here we focus only on those that have been already searched for at the LHC, which we collect in Table~\ref{LHCexp}.
We notice that similar searches considering the existence of right-handed currents have also been performed~\cite{ATLAS:2011izm,ATLAS:2012ak,CMS:2012zv,CMS:2014nrz,CMS:2018agk,ATLAS:2018dcj,CMS:2018iye,ATLAS:2019isd}, and could in principle be recast to our setup. 
Nevertheless, one would naively expect lower sensitivities, as they are optimized for heavy right-handed gauge bosons.

Most of the LHC searches focused on the {\it smoking gun} signature for Majorana neutrinos, the same sign (SS) dilepton final state:
\begin{equation}
p p \to \ell_\alpha^\pm N \to \ell_\alpha^\pm \ell_\beta^\pm + n j\,.
\end{equation}
Here, the lepton pair is accompanied by at least two jets (see Fig.~\ref{DiagHNLatLHC}), unless $M_N$ is much lighter or much heavier than $M_W$, which leads to boosted objects and collimated jets that are reconstructed as a single one.

Being a LNV process, the SS dilepton does not suffer from severe SM backgrounds.
Unfortunately, current collider searches are sensitive only to relatively large mixings between the HNL and the active neutrinos, too large to explain the masses of the light neutrinos unless a symmetry protected scenario is invoked.
More specifically, this symmetry is an approximated conservation of lepton number~\cite{Moffat:2017feq}, which also suppressed the expected LNV signal from HNLs (see however Refs.~\cite{Antusch:2017ebe,Drewes:2019byd,Fernandez-Martinez:2022gsu}).

From this point of view, searching for opposite sign (OS) dileptons, as done by LHCb~\cite{LHCb:2020wxx}, seems more relevant to explore theoretically motivated scenarios.
The drawback is the large amount of background from $Z\to\ell^+\ell^-$ decays, which reduces the sensitivity. 
A possible alternative would be focusing on LFV channels to reduce backgrounds~\cite{Arganda:2015ija,Antusch:2018bgr}.

Yet another alternative considers the fully leptonic process
 \begin{equation}
p p \to \ell_\alpha^\pm N \to \ell_\alpha^\pm \ell_\beta^\pm \ell_\gamma^\mp \nu\,.
\end{equation}
This channel has a trilepton signature, rather clean in a hadronic collider, nevertheless it also has a source of MET, which might spoil the complete reconstruction of $M_N$.
The trilepton channel offers the possibility to search for both LNV and LNC signals, however most of the experimental analyses still focus only on the LNV channels as to reduce backgrounds, again from $Z\to\ell^+\ell^-$.
For example, ATLAS searched~\cite{ATLAS:2019kpx} for $e^\pm e^\pm \mu^\mp$ and $\mu^\pm \mu^\pm e^\mp$ channels\footnote{Having an undetected (anti)neutrino, it is not always possible to define a LNV or LNC process unambiguously. Assuming the presence of a HNL that mixes only to electrons or to muons, as ATLAS did, the $e^\pm e^\pm \mu^\mp$ and $\mu^\pm \mu^\pm e^\mp$ channels are originated only from LNV processes. However, this is not anymore true if the HNL mixes to both flavours~\cite{Tastet:2021vwp}.}, but not for $e^\pm e^\mp \mu^\pm$ and $\mu^\pm \mu^\mp e^\pm$.
CMS did something similar for light HNLs, although they also included channels with OS but same flavour lepton pairs in the heavy HNL regime, removing only those events with lepton pairs compatible with a decay of a $Z$ boson~\cite{CMS:2018jxx}.
As stated before for the SS dilepton channel, searching for HNLs without assuming their Majorana nature will be helpful to probe scenarios compatible with neutrino oscillation data, and thus with potentially suppressed LNV signals. 

\begin{figure}[t!]
\begin{center}
\includegraphics[width=\textwidth]{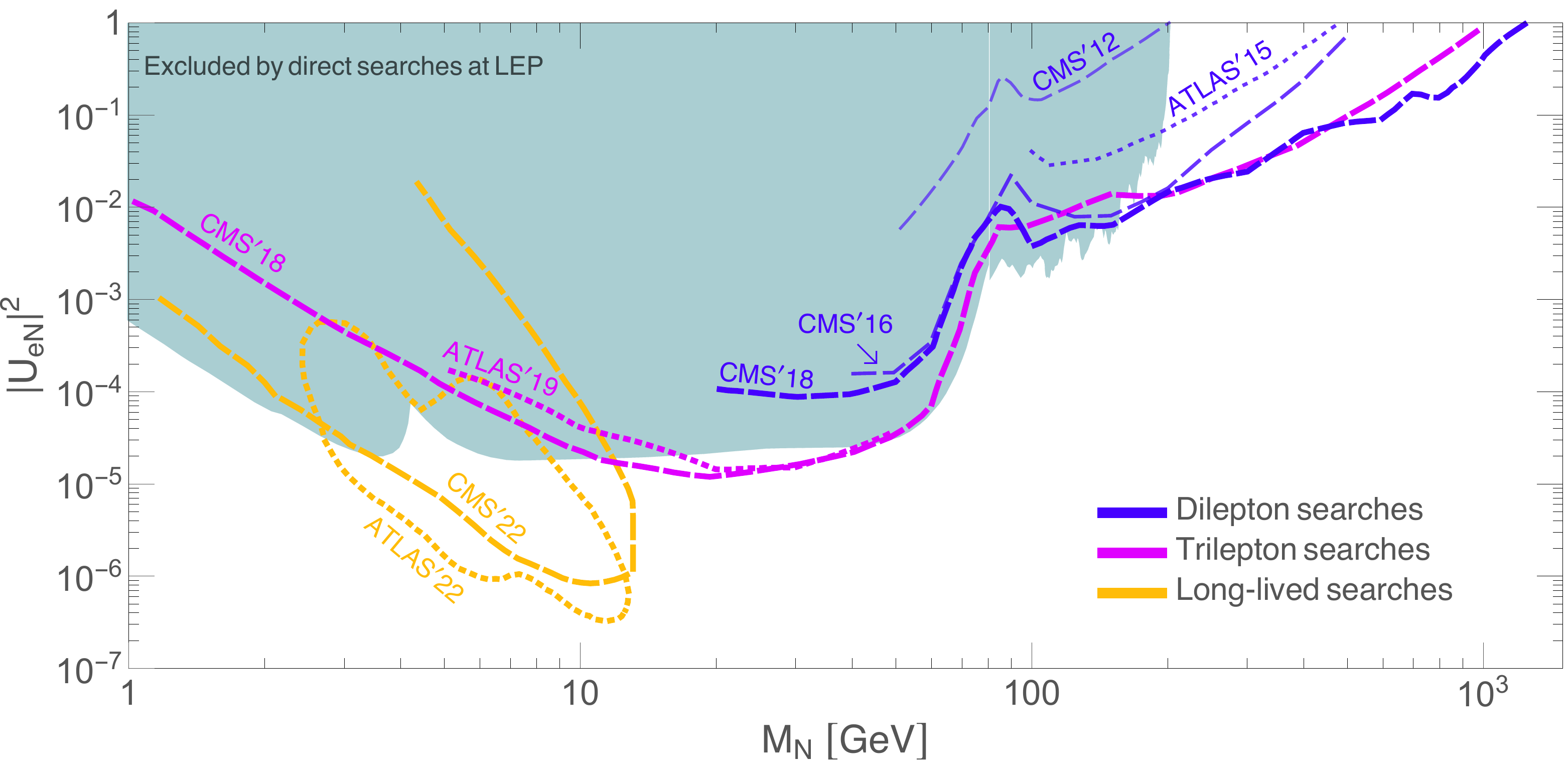}
\includegraphics[width=\textwidth]{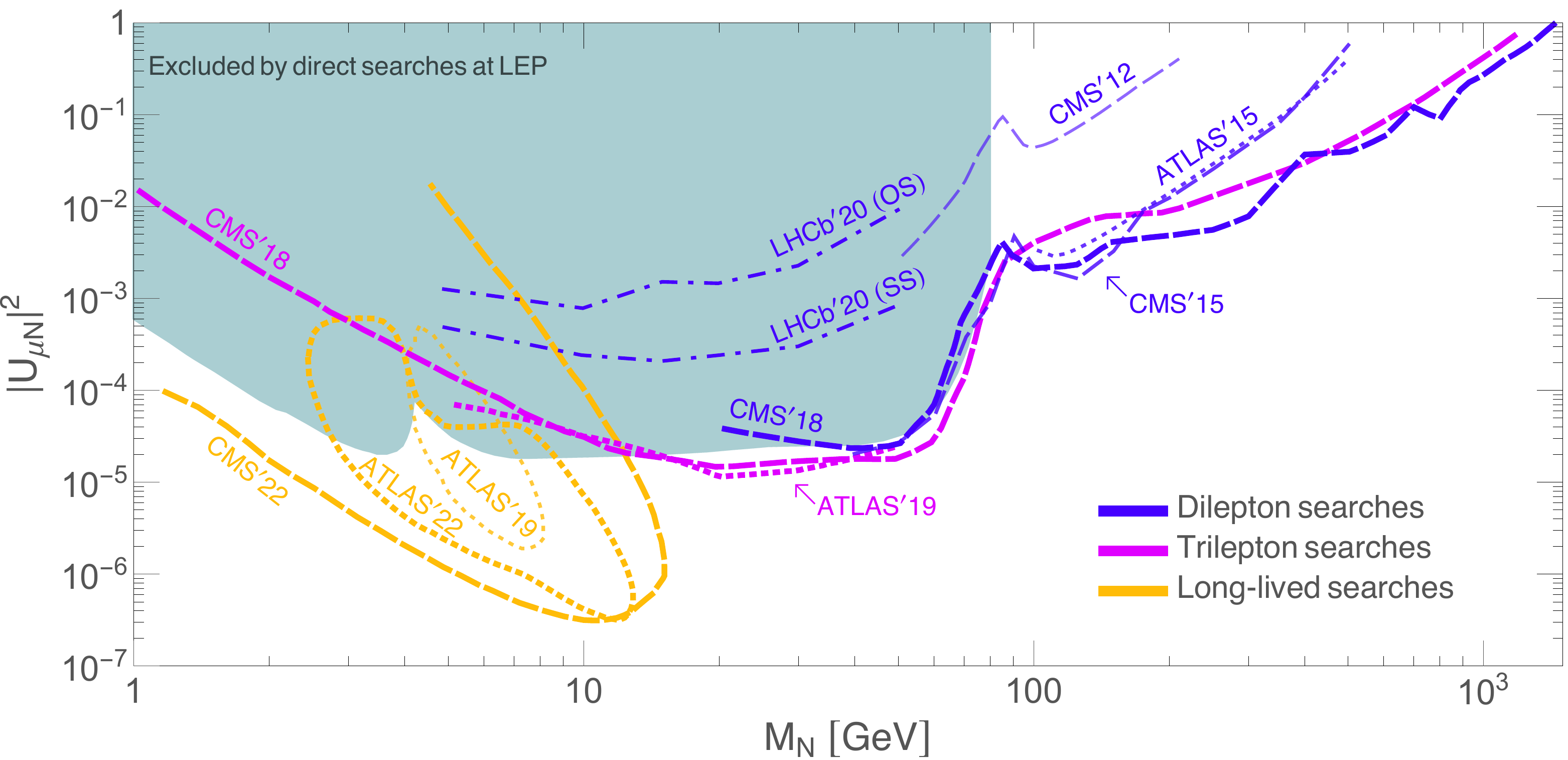}
\caption{Summary of direct HNL searches performed at the LHC so far, either by CMS (dashed), ATLAS (dotted) or LHCb (dot-dashed), and group by colors for different kind of searches as given in Table~\ref{LHCexp}.
In the upper (lower) panel a single mixing scenario to electrons (muons) is assumed.
Shadowed area cover the area excluded by direct searches at LEP.
Notice that below 2~GeV and above (approx.) 100~GeV, bounds from meson decays and from non-unitarity of the lepton mixing~\cite{Fernandez-Martinez:2016lgt} dominate respectively over current LHC bounds, although we do not show them explicitly for easier reading of the collider results.
}\label{BoundsColliders}
\end{center}
\end{figure}

Finally, it is important to stress that improving the experimental sensitivities to smaller values of mixings implies exploring HNL with longer lifetimes, which can travel macroscopic distances before decaying.
Such long-lived HNL would avoid the searches mentioned so far, as they all assumed prompt decaying HNLs, and therefore we need a dedicated search for this kind of topologies. 
Recently, both ATLAS~\cite{ATLAS:2022atq} and CMS~\cite{CMS:2022fut} have searched for displaced vertex signatures for light HNLs, setting the strongest constraints for GeV masses up to 20~GeV.
This can be seen in Fig.~\ref{BoundsColliders}, where we summarize all the relevant LHC constraints explained in this section.

\begin{table}[t!]
\begin{center}
\small{
\renewcommand{\arraystretch}{1.1}
\begin{tabular}{|c|c|l|c|c|c|}
\hline
Channel & Lepton Flavour & Experiment & $\sqrt s$~[TeV] & $\mathcal L~[{\rm fb}^{-1}]$ & $M_N$~[GeV]  \\
\hline
\multirow{6}*{\minitab[c]{
prompt SS dilepton \\[1ex]
$p p \to \ell_\alpha^\pm N \to \ell_\alpha^\pm \ell_\beta^\pm + n j$}}
& $ee / \mu\mu$ & CMS'12~\cite{CMS:2012wqj} & 7 & 4.98 & (50, 210)\\
& $\mu\mu$ & CMS'15~\cite{CMS:2015qur} & 8 & 19.7 & (40, 500)\\
& $ee/e\mu$ & CMS'16~\cite{CMS:2016aro} & 8 & 19.7 & (40, 500)\\
& $ee/\mu\mu$ & ATLAS'15~\cite{ATLAS:2015gtp} & 8 & 20.3 & (100, 500)\\
& $ee / e\mu /\mu\mu$ & CMS'18~\cite{CMS:2018jxx} & 13 & 35.9 & (20, 1600)\\
& $\mu\mu$ & LHCb'20~\cite{LHCb:2020wxx} & 7-8 & 3.0 & (5, 50)\\[1ex]
\hline 
&&&&&\\[-.4cm]
\minitab[c]{
prompt OS dilepton \\
$p p \to \ell_\alpha^\pm N \to \ell_\alpha^\pm \ell_\beta^\mp + n j$}
& $\mu\mu$ & LHCb'20~\cite{LHCb:2020wxx} & 7-8 & 3.0 & (5, 50)\\[2ex]
\hline 
&&&&&\\[-.4cm]
\multirow{2}*{\minitab[c]{
Prompt trilepton \\
$p p \to \ell_\alpha^\pm N \to \ell_\alpha^\pm \ell_\beta^\pm \ell_\gamma^\mp \nu$}}
& $eee+ee\mu / \mu\mu\mu+\mu\mu e$ & CMS'18~\cite{CMS:2018iaf} & 13 & 35.9 & (1, 1200)\\
& $ee\mu /\mu\mu e$ & ATLAS'19~\cite{ATLAS:2019kpx} & 13 & 36.1 & (5, 50)\\[1ex]
\hline 
&&&&&\\[-.4cm]
\multirow{3}*{\minitab[c]{
Displaced trilepton \\[1ex]
$p p \to \ell_\alpha N,~ N \to \ell_\beta \ell_\gamma \nu$}}
& $\mu - e\mu/ \mu- \mu\mu$ & ATLAS'19~\cite{ATLAS:2019kpx} & 13 & 32.9 & (4.5, 10)\\
& 6 combinations of $e,\mu$ & ATLAS'22~\cite{ATLAS:2022atq} & 13 & 139 & (3, 15)\\
& 6 combinations of $e,\mu$  & CMS'22~\cite{CMS:2022fut} & 13 & 138 & (1, 20)\\[1ex]
\hline

\end{tabular}
\caption{HNL searches at the LHC, classified according to the type of signal searched for. 
OS/SS are for opposite/same sign for the charges of final leptons and $nj$ for a number $n$ of final jets.} \label{LHCexp} 
}
\end{center}
\end{table}

\subsection*{Searches at lepton colliders}

Despite the fact that our current most powerful high-energy collider is a hadronic one, it is  important to stress  that lepton colliders are extremely relevant for HNL searches. 
Not only due to the impressive sensitivities expected at future leptonic colliders such as the FCCee~\cite{Blondel:2014bra}, but also because HNL searches at LEP still provide the most relevant limits for some $M_N$ hypotheses.

The great advantage of a leptonic collider is its clean environment, in contrast with the  hadronic ones.
In the case of LEP, they combined this cleanliness with the huge amount of $Z$ bosons they collected to search for HNLs produced in $Z\to \nu N$. 
Moreover, they considered both visibles and semi-invisible HNL decays, such as monojet final states~\cite{DELPHI:1996qcc}:
\begin{equation}
e^+ e^- \to \nu N \to \nu \nu q \bar q\,,
\end{equation}
with the $q\bar q$ pair clustered as a single jet due to the large HNL boost (efficient for $M_N\lesssim 30$~GeV).
For heavier masses, $M_N\in (30,80)$~GeV, the signature was composed by two jets with or without a charged lepton.
Such a search would be very challenging at a hadronic collider, however it has the advantage of being sensitive to all flavours, including the mixing to the  $\tau$ lepton, not explored so far by  LHC searches.
The DELPHI resutls~\cite{DELPHI:1996qcc}, derived for both long-lived and prompt light HNLs, were not improved (for mixings to $e$ and $\mu$ flavours) by LHC until very recently, and still dominate for some mass ranges (cf. Fig.~\ref{BoundsColliders}).

Additionally, the L3 collaboration explored the heavy HNL regime by considering their production via the t-channel $W$ diagram~\cite{L3:2001zfe}.
This process dominates the heavy HNL production at a $e^+ e^-$ collider running above the $Z$ pole, however it is sensitive only to mixings to electrons. 
The results by L3 still provide the strongest limits for masses between 100 and 200~GeV.

Despite the great effort in the search for HNLs by both LEP and LHC, it is important to analyze their implications for realistic models introducing and motivating the existence of HNLs. 
A common feature of all these searches is the assumption of a simplified scenario, most of the time consisting on a single HNL mixing to a single lepton flavour, which is not the standard hypothesis one would use from the theory side.
To our knowledge, the only exceptions to these simplifications are provided by the recent ATLAS search for long-lived HNLs that also considered a minimal but realistic 2HNL scenario~\cite{ATLAS:2022atq}, and CMS searches for SS $e\mu$ final states~\cite{CMS:2016aro,CMS:2018jxx}, although still neglecting the mixing to taus.
In the  two following sections, we discuss the importance of going beyond these simplified scenarios to explore more realistic scenarios.

\section{Beyond the single mixing assumption}\label{sec:flavour}

As explained in the previous section, most of the LHC analyses are done assuming the existence of just one HNL that mixes to a single flavour, which we referred to as the single mixing scenario.
In this section, we consider deviations from this simplified hypothesis and discuss their implications for reinterpreting the LHC bounds summarized in Fig.~\ref{BoundsColliders}.
In particular, we focus only on prompt searches, while the implications for long-lived HNL were discussed in, for instance,  Ref.~\cite{Abada:2018sfh}.

For simplicity, we still consider the presence of a single HNL (we will discuss deviations from this hypothesis in the next section), however we open the room for generic mixing patterns.
Moreover, we will follow a bottom-up approach where the SM is extended by {\it ad-hoc} masses for the 3 active neutrinos, as required by oscillation phenomena, and by the presence of the additional HNL $N$.
This framework is useful to study, in particular, the collider phenomenology of HNLs without assuming any specific underlying model or mechanism of light neutrino mass and leptonic mixing generation\footnote{Notice however that reproducing oscillation data in a given framework may introduce relations between the HNL mass and mixings, shrinking the parameter space we will consider.}.

In such a framework, the lepton mixing matrix is thus enlarged to a $4\times4$ unitary matrix 
\begin{equation}
U_\nu = \left(\begin{array}{cc}
U_{\nu\nu}^{3\times3} & U_{\nu N}^{3\times1} \\
U_{N\nu}^{1\times3} & U_{NN}^{1\times1}
\end{array}\right)\,,
\end{equation}
so the would-be-PMNS matrix $U_{\nu\nu}$ is no longer a unitary matrix, a feature which is indeed used to constrain these models~\cite{Fernandez-Martinez:2016lgt}, and the fourth column contains the HNL mixings to each flavour:
\begin{equation}
U_{\nu N}^T = \big(U_{e N}, U_{\mu N}, U_{\tau N} \big)\,.
\end{equation}
For our discussion, it is interesting to parametrize this column as
\begin{equation}\label{eq:usquardef}
U_{\nu N}^T = \sqrt{U^2}\, \big(\varepsilon_e,\varepsilon_\mu,\varepsilon_\tau \big)\,, 
\end{equation}
where $U^2$ represents the total (squared) mixing of the HNL and the $\varepsilon_\alpha$ its flavour strengths, with $|\varepsilon_e|^2 + |\varepsilon_\mu|^2 + |\varepsilon_\tau|^2 = 1$.
Notice that this framework is precisely the one considered by most LHC analyses, the only difference being that they simplify it by setting the a priori non-relevant mixings to zero. 
Here, we are interested in knowing how these bounds need to be modified in a generic mixing pattern scenario.

The reason why we expect the bounds to be modified is twofold.
The first reason is due to the importance of the HNL decay width, which depends on every mixing $U_{\alpha N}$, and which plays a major role in the resonant searches (on-shell produced HNL) we are interested in.
This means that the final cross sections will depend on all of the mixings, even on those flavours that are not explicitly present in the charged leptons involved in the process. 
The second reason is that for some channels, considering generic mixings could open new contributing diagrams, which could modify the distributions and thus the efficiencies of the searches as  discussed thoroughly in Ref.~\cite{Tastet:2021vwp}. 
Complete expressions for the computation of the total decay width $\Gamma_N$ can be found for instance in Ref.~\cite{Bondarenko:2018ptm}, however for our purposes we parameterize it as
\begin{equation}
\Gamma_N = |U_{eN}|^2\, \Gamma_{N}^e + |U_{\mu N}|^2\, \Gamma_{N}^\mu +|U_{\tau N}|^2\, \Gamma_{N}^\tau\,,
\end{equation}
where $\Gamma_N^{\alpha}$ stands for the sum of partial decay widths depending on the mixing $U_{\alpha N}$, after factorizing the $|U_{\alpha N}|^2$ dependence itself.
Thus, $\Gamma_N^\alpha$ are independent of the mixings (at leading order) and depend only on the HNL mass.
Moreover, when the HNL is heavy enough so that we can neglect charged lepton masses, we get $\Gamma_N^{e}\simeq \Gamma_N^{\mu} \simeq \Gamma_N^{\tau}$ and thus
\begin{equation}
\Gamma_N\propto \sum_\alpha \big|U_{\alpha N}\big|^2 = U^2\,.
\end{equation}

With this discussion in mind, 
we can now study how the different processes displayed in Table~\ref{LHCexp} depend on the HNL mixings.
Let us start focusing on the dilepton channels, the most straighforward case as we only need to track the effect of the HNL total  decay width.

In the narrow width approximation, the processes with SS and same flavour dileptons can be factorized in the production of the HNL together with a charged lepton, times its subsequent decay to the same lepton plus jets. 
The first part depends only on the mixing to the flavour of that lepton, however the second one involves all the mixings due to the HNL decay width. 
More explicitly, we have
\begin{equation}\label{eq:sigmadilepton}
\sigma (pp\to\ell_\alpha^\pm N \to \ell_\alpha^\pm\ell_\alpha^\pm +nj)
\propto |U_{\alpha N}|^2\, {\rm BR}(N\to\ell_\alpha^\pm jj) 
\propto \frac{|U_{\alpha N}|^4}{\Gamma_N}\,,
\end{equation}
or, assuming a heavy enough HNL, 
\begin{equation}
\sigma (pp\to\ell_\alpha^\pm N \to \ell_\alpha^\pm\ell_\alpha^\pm +nj) 
\propto U^2\, |\varepsilon_\alpha|^4\,.
\end{equation}
Then, we clearly see that those bounds obtained in the single mixing benchmark ($\varepsilon_\alpha=1$) will be relaxed in a general flavour scenario with a fixed $U^2$, since in general we will have $|\varepsilon_\alpha|^2\leq1$.
This is actually the expected behaviour, as switching on other mixings opens for new decay channels, so not every produced HNL will decay to the final state we are searching for.

\begin{table}[t!]
\begin{center}
\small{
\renewcommand{\arraystretch}{1.1}
\begin{tabular}{|c|c|c|}
\hline
\multirow{2}*{process (prompt)} & \multicolumn{2}{c|}{Relevant parameters (Majorana HNL)}\\
\cline{2-3}
& approx. & complete dependence \\
\hline&&\\[-2ex]
$pp\to \ell_\alpha^\pm \ell_\alpha^\pm + nj$ & 
$U^2\, |\varepsilon_\alpha|^4$ &
$\big|U_{\alpha N}\big|^2\, {\rm BR}(N\to\ell_\alpha^\pm jj)$  
\\[1ex]
$pp\to \ell_\alpha^\pm \ell_\beta^\pm + nj$ &
$U^2\, |\varepsilon_\alpha|^2\, |\varepsilon_\beta|^2$ &  
$|U_{\alpha N}|^2\, {\rm BR}(N\to\ell_\beta^\pm jj) + |U_{\beta N}|^2\, {\rm BR}(N\to\ell_\alpha^\pm jj)$  \\[2ex]
\hline&&\\[-4ex]\hline&&\\[-1ex]
$pp\to \ell_\alpha^+\ell_\alpha^+\ell_\alpha^- + \slashed{E}_T$ &
$U^2\, |\varepsilon_\alpha|^4$ & 
$|U_{\alpha N}|^2\, {\rm BR}(N\to\ell_\alpha^+\ell_\alpha^-\bar\nu_\alpha) +|U_{\alpha N}|^2\, {\rm BR}(N\to\ell_\alpha^-\ell_\alpha^+\nu_\alpha)$ \\[1ex]
$pp\to \ell_\alpha^+\ell_\alpha^+\ell_\beta^- +\slashed{E}_T$ &
$U^2\, |\varepsilon_\alpha|^2\, (|\varepsilon_\alpha|^2+|\varepsilon_\beta|^2)$ &
$|U_{\alpha N}|^2\, {\rm BR}(N\to\ell_\alpha^+\ell_\beta^-\bar\nu_\beta)
+  |U_{\alpha N}|^2\, {\rm BR}(N\to\ell_\beta^-\ell_\alpha^+\nu_\alpha)$ \\[1ex]
$pp\to \ell_\alpha^+\ell_\alpha^-\ell_\beta^++\slashed{E}_T$ &
$U^2\, |\varepsilon_\alpha|^2\,(|\varepsilon_\alpha|^2+3|\varepsilon_\beta|^2)$ &
$|U_{\alpha N}|^2\, {\rm BR}(N\to\ell_\alpha^-\ell_\beta^+\nu_{\phantom \alpha})
+  |U_{\beta N}|^2\, {\rm BR}(N\to\ell_\alpha^-\ell_\alpha^+\nu_{\phantom \alpha}) $ \\[1ex]
$pp\to \ell_\alpha^+\ell_\beta^+\ell_\gamma^-+\slashed{E}_T$ &
$U^2\, {\scriptscriptstyle\displaystyle\sum_{\scriptstyle i=\alpha,\beta}}|\varepsilon_i|^2\,\big(1- |\varepsilon_i|^2\big)$ &
$|U_{\alpha N}|^2\, {\rm BR}(N\to\ell_\beta^+\ell_\gamma^-\nu_{\phantom \alpha})
+  |U_{\beta N}|^2\, {\rm BR}(N\to\ell_\alpha^+\ell_\gamma^-\nu_{\phantom \alpha})$\\[1ex]
\hline
\end{tabular}
\caption{Summary table for generic flavour dependences of dilpeton and trilepton channels at the LHC assuming a single Majorana HNL with generic mixing patterns.
Flavour indices are to be understood as different, {\it i.e.} $\alpha\neq\beta\neq\gamma$.
For trileptons, we are neglecting the effects of differential distributions, as discussed in the text.
}\label{Tab:recastingMixingMajorana}
}
\end{center}
\end{table}

We can repeat the exercise for the different flavour SS dilepton processes.
Obviously, the minimal setup in this case requires to have two non-zero mixings, which leads to two diagrams that in principle interfere. 
Nevertheless, using the narrow width approximation, we can see that both diagrams cannot resonate at the same time, so we can neglect the interference and add both processes incoherently:
\begin{align}
\sigma (pp\to \ell_\alpha^\pm\ell_\beta^\pm +nj)
&\propto \Big(|U_{\alpha N}|^2\, {\rm BR}(N\to\ell_\beta^\pm jj) +|U_{\beta N}|^2\, {\rm BR}(N\to\ell_\alpha^\pm jj) \Big)\nonumber\\
&\propto \frac{|U_{\alpha N}|^2|U_{\beta N}|^2}{\Gamma_N}
\propto U^2\, |\varepsilon_\alpha|^2\, |\varepsilon_\beta|^2\,.
\end{align}
The case of the trilepton channels can be more involved, mainly because we  
cannot know the lepton number and the flavour carried by the missing (anti)neutrino.
Let us consider first the case of same flavour trileptons.
There are two contributing diagrams, one with a neutrino and one with an antineutrino, which we can add incoherently\footnote{If light neutrinos are of Majorana nature, then there is an interference term, which is however negligible as it is proportional to light neutrino masses.}.
Then, considering for simplicity  a $W^+$ Drell-Yan channel, we have
\begin{equation}
\sigma (pp\to \ell_\alpha^+\ell_\alpha^+\ell_\alpha^- +\slashed{E}_T)
\propto \Big(|U_{\alpha N}|^2\, {\rm BR}(N\to\ell_\alpha^+\ell_\alpha^-\bar\nu_\alpha) +|U_{\alpha N}|^2\, {\rm BR}(N\to\ell_\alpha^-\ell_\alpha^+\nu_\alpha)\Big)\,.
\end{equation}
Notice that the first contribution is mediated by the Majorana nature of the HNL, while the second one is of Dirac type.
In principle, both contributions are identical when integrated over the full phase space. 
However, different spin-correlations induce different angular distributions, which might translate into different acceptances under a given experimental analysis.
Still, both channels have the same flavour dependence, so we can write, 
\begin{equation}
\sigma (pp\to \ell_\alpha^+\ell_\alpha^+\ell_\alpha^- +\slashed{E}_T)
\propto |U_{\alpha N}|^2\, {\rm BR}(N\to\ell_\alpha\ell_\alpha\nu_\alpha) 
\propto \frac{|U_{\alpha N}|^4}{\Gamma_N}\propto U^2\, |\varepsilon_\alpha|^4\,.
\end{equation}

When the trilepton signal involves leptons of two different flavours, we need to consider two subcases: the same-sign same-flavour (SSSF) and opposite-sign same-flavour (OSSF).
These channels are trickier because in a generic flavour pattern there are new diagrams not present in the single mixing scenario.
For instance, in the case of the SSSF, we have two types of contributions:
\begin{equation}
\begin{array}{ll}
\text{\rm Majorana-like:}&  pp\to \ell^+_\alpha N\,, N\to \ell^+_\alpha \ell^-_\beta \bar \nu_\beta\,,\\
\text{\rm Dirac-like:} &  pp \to \ell_\alpha^+ N\,, N\to \ell_\beta^-\ell_\alpha^+\nu_\alpha\,.
\end{array}
\end{equation}
It is clear that the first process requires of a Majorana HNL, while the second one needs mixings to both flavours.
As before, the interference is negligible, so we have
\begin{equation}
\sigma (pp\to \ell_\alpha^+\ell_\alpha^+\ell_\beta^- +\slashed{E}_T)
\propto \Big(|U_{\alpha N}|^2\, {\rm BR}(N\to\ell_\alpha^+\ell_\beta^-\bar\nu_\beta)
+  |U_{\alpha N}|^2\, {\rm BR}(N\to\ell_\beta^-\ell_\alpha^+\nu_\alpha)\Big)\,.
\end{equation}
Due to the missing (anti)neutrino, both processes are almost identical at the LHC, with the only difference coming again from the different distributions and acceptances of the experimental analysis.
This was studied in detail in Ref.~\cite{Tastet:2021vwp} for the case of light HNLs. 
Nevertheless, in order to get a first rough estimate, we can neglect these differences and write
\begin{equation}
\sigma (pp\to \ell_\alpha^+\ell_\alpha^+\ell_\beta^- +\slashed{E}_T)
\propto  |U_{\alpha N}|^2\, \frac{|U_{\alpha N}|^2 + |U_{\beta N}|^2}{\Gamma_N}
\propto U^2\, |\varepsilon_\alpha|^2\, (|\varepsilon_\alpha|^2+|\varepsilon_\beta|^2)\,.
\end{equation}
The case of OSSF is similar, although now the roles of Majorana and Dirac HNLs are flipped:
\begin{equation}
\begin{array}{ll}
\text{\rm Majorana-like:}&  pp\to \ell^+_\alpha N\,, N\to \ell^+_\beta \ell^-_\alpha \bar \nu_\alpha\,,\\
\text{\rm Dirac-like:} &  pp \to \ell_\alpha^+ N\,, N\to \ell_\alpha^-\ell_\beta^+\nu_\beta\,.
\end{array}
\end{equation}
Moreover, since we are working in the prompt HNL regime, we can also have $pp\to \ell_\beta^+ N, N\to \ell_\alpha^+\ell_\alpha^- \overset{\textbf{\fontsize{1pt}{2pt}\selectfont(--)}}{\nu}_\alpha $.
This means that a proper recasting of this kind of signals would require to compute the efficiencies for all these diagrams. 
If, for the shake of this discussion, we neglect these effects again, we get:
\begin{align}
\sigma (pp\to \ell_\alpha^+\ell_\alpha^-\ell_\beta^++\slashed{E}_T)
&\propto \Big(|U_{\alpha N}|^2\, {\rm BR}(N\to\ell_\alpha^-\ell_\beta^+\nu_\beta)
+  |U_{\alpha N}|^2\, {\rm BR}(N\to\ell_\beta^+\ell_\alpha^-\bar\nu_\alpha)\nonumber\\
&~~+\, |U_{\beta N}|^2\, {\rm BR}(N\to\ell_\alpha^-\ell_\alpha^+\nu_\alpha)
+  |U_{\beta N}|^2\, {\rm BR}(N\to\ell_\alpha^+\ell_\alpha^-\bar\nu_\alpha)\Big )\nonumber\\
&\propto |U_{\alpha N}|^2\, \frac{|U_{\alpha N}|^2 + 3 |U_{\beta N}|^2}{\Gamma_N}
\propto U^2\, |\varepsilon_\alpha|^2\,(|\varepsilon_\alpha|^2+3|\varepsilon_\beta|^2)\,.
\end{align}

Finally, and even if no LHC searches have been performed so far, we can also consider the case with 3 different flavours. Following the same steps, we get
\begin{align}
\sigma (pp\to \ell_\alpha^+\ell_\beta^+\ell_\gamma^-+\slashed{E}_T)
&\propto \Big(|U_{\alpha N}|^2\, {\rm BR}(N\to\ell_\beta^+\ell_\gamma^-\bar\nu_\gamma)
+\,  |U_{\alpha N}|^2\, {\rm BR}(N\to\ell_\gamma^-\ell_\beta^+\nu_\beta)\nonumber\\
&~~+\, |U_{\beta N}|^2\, {\rm BR}(N\to\ell_\alpha^+\ell_\gamma^-\bar\nu_\alpha)
+  |U_{\beta N}|^2\, {\rm BR}(N\to\ell_\alpha^-\ell_\alpha^+\nu_\alpha)\Big) \nonumber\\
&
\propto U^2\, \Big\{|\varepsilon_\alpha|^2\,\big(1- |\varepsilon_\alpha|^2\big)+|\varepsilon_\beta|^2\,\big(1- |\varepsilon_\beta|^2\big)\Big\}\,.
\end{align}

\begin{table}[t!]
\begin{center}
\small{
\renewcommand{\arraystretch}{1.1}
\begin{tabular}{|c|c|c|}
\hline
\multirow{2}*{process (prompt)} & \multicolumn{2}{c|}{Relevant parameters (Dirac HNL)}\\
\cline{2-3}
& approx. & complete dependence \\
\hline&&\\[-2ex]
$pp\to \ell_\alpha^\pm \ell_\alpha^\mp + nj$ & 
$U^2\, |\varepsilon_\alpha|^4$ &
$\big|U_{\alpha N}\big|^2\, {\rm BR}(N\to\ell_\alpha^\pm jj)$  
\\[1ex]
$pp\to \ell_\alpha^\pm \ell_\beta^\mp + nj$ &
$U^2\, |\varepsilon_\alpha|^2\, |\varepsilon_\beta|^2$ &  
$|U_{\alpha N}|^2\, {\rm BR}(N\to\ell_\beta^\pm jj) + |U_{\beta N}|^2\, {\rm BR}(N\to\ell_\alpha^\pm jj)$ \\[2ex]
\hline&&\\[-4ex]\hline&&\\[-1ex]
$pp\to \ell_\alpha^+\ell_\alpha^+\ell_\alpha^- + \slashed{E}_T$ &
$U^2\, |\varepsilon_\alpha|^4$ & 
$|U_{\alpha N}|^2\, {\rm BR}(N\to\ell_\alpha^-\ell_\alpha^+\nu_\alpha)$ \\[1ex]
$pp\to \ell_\alpha^+\ell_\alpha^+\ell_\beta^- +\slashed{E}_T$ &
$U^2\, |\varepsilon_\alpha|^2|\varepsilon_\beta|^2$ &
$|U_{\alpha N}|^2\, {\rm BR}(N\to\ell_\beta^-\ell_\alpha^+\nu_\alpha)$ \\[1ex]
$pp\to \ell_\alpha^+\ell_\alpha^-\ell_\beta^++\slashed{E}_T$ &
$U^2\, |\varepsilon_\alpha|^2\,(|\varepsilon_\alpha|^2+|\varepsilon_\beta|^2)$ &
$|U_{\alpha N}|^2\, {\rm BR}(N\to\ell_\alpha^-\ell_\beta^+\nu_{\beta})
+  |U_{\beta N}|^2\, {\rm BR}(N\to\ell_\alpha^-\ell_\alpha^+\nu_{\alpha}) $ \\[1ex]
$pp\to \ell_\alpha^+\ell_\beta^+\ell_\gamma^-+\slashed{E}_T$ &
$U^2\,|\varepsilon_\gamma|^2\,\big(1- |\varepsilon_\gamma|^2\big)$ &
$|U_{\alpha N}|^2\, {\rm BR}(N\to\ell_\gamma^-\ell_\beta^+\nu_{\beta})
+  |U_{\beta N}|^2\, {\rm BR}(N\to\ell_\gamma^-\ell_\alpha^+\nu_\alpha)$\,\\[1ex]
\hline
\end{tabular}
\caption{Same as Table~\ref{Tab:recastingMixingMajorana}, but for Dirac HNL. 
We give the OS dileptons in this case, since the SS are not sensitive to Dirac HNLs.
}\label{Tab:recastingMixingDirac}
}
\end{center}
\end{table}

We summarize our discussion in Table~\ref{Tab:recastingMixingMajorana}.
Here, we assume a Majorana HNL, although a similar table can be easily obtained for Dirac HNL by just switching off the LNV channels we discussed above. 
This is done in Table~\ref{Tab:recastingMixingDirac}.
Notice how some channels that were designed for Majorana HNLs are also sensitive to Dirac HNLs in the case of having generic mixing patterns, as it was already discussed in Ref.~\cite{Tastet:2021vwp}.

These tables are to be compared with the minimal mixing scenario where all  of the processes  scale as $|U_{\alpha N}|^2$, or $|U_{\alpha N}|^2|U_{\beta N}|^2/(|U_{\alpha N}|^2+|U_{\beta N}|^2)$ for $\alpha\neq\beta$.
However, we see that in general each process is sensitive to a different combinations of mixing strengths. 
This means that, in order to generalize the bounds to a generic pattern, it is better to set bounds on the quantity on the last columns of Tables~\ref{Tab:recastingMixingMajorana} and \ref{Tab:recastingMixingDirac}, since then we only need to recompute the new BRs for each mixing hypothesis.

While there are some experimental results also presenting  (in the single mixing assumption) bounds in the ($M_N, |U_{\alpha N}|^2\times$BR) plane, most of the results are given directly in the ($M_N, |U_{\alpha N}|^2)$ one.
Translating the latter to the former is straightforward in most of the cases, since the experimental collaborations usually assume a constant\footnote{This is a well-justified approximation, but still an approximation that could be avoided by setting bounds directly on $|U_{\alpha N}|^2\times$BR.} BR for channel under study, although not always specifying the precise value they used.
Nevertheless, this recasting to $|U_{\alpha N}|^2\times$BR is not possible when the experimental results on $|U_{\alpha N}|^2$ are presented after combining different channels.
This is the case for instance for the latest CMS searches for trileptons~\cite{CMS:2018iaf}, where they combined channels like $e^\pm e^\pm e^\mp$ and $e^\pm e^\pm \mu^\mp$ (see Table~\ref{LHCexp}).
While in the single mixing scenario both channels depend only on $|U_{e N}|^2$,  they have a different dependence in the case of a generic mixing scenario (see Table~\ref{Tab:recastingMixingMajorana}), and thus it is not easy to recast the obtained bounds without a dedicated analysis.
For this reason, together with the potential efficiency differences discussed above, we will focus the rest of our discussion only on the dilepton channels.

\begin{figure}[t!]
\centerline{\includegraphics[width=.43\linewidth]{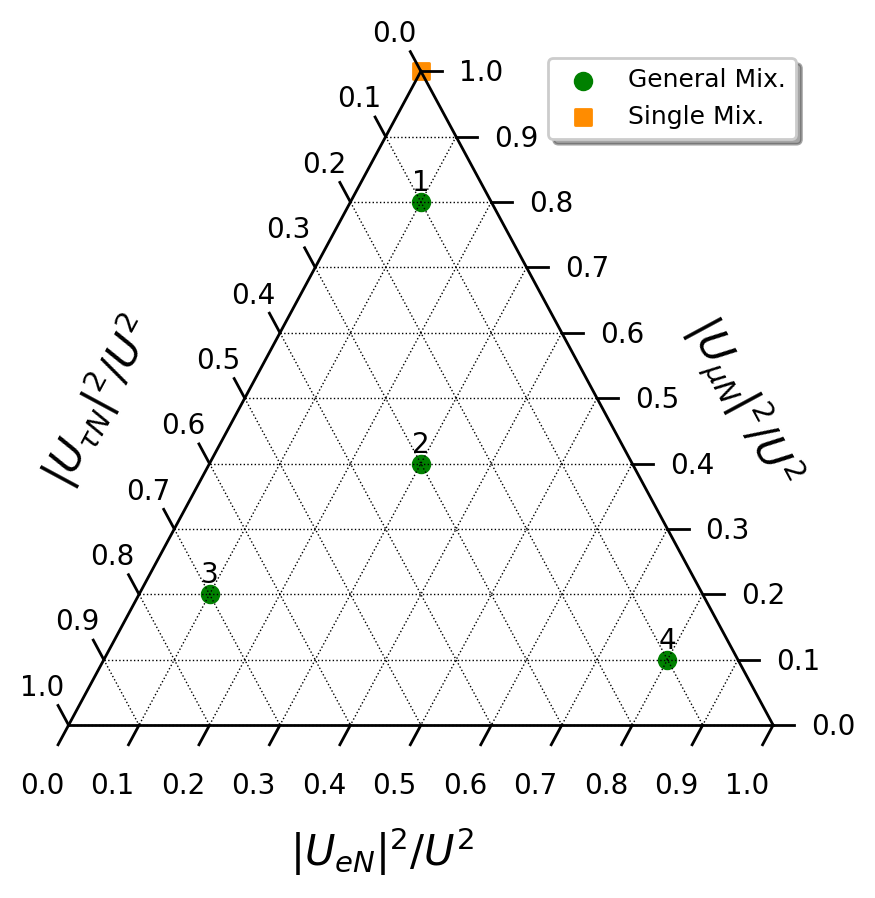}  \raisebox{3.4ex}{ {\includegraphics[ width=.58\linewidth]{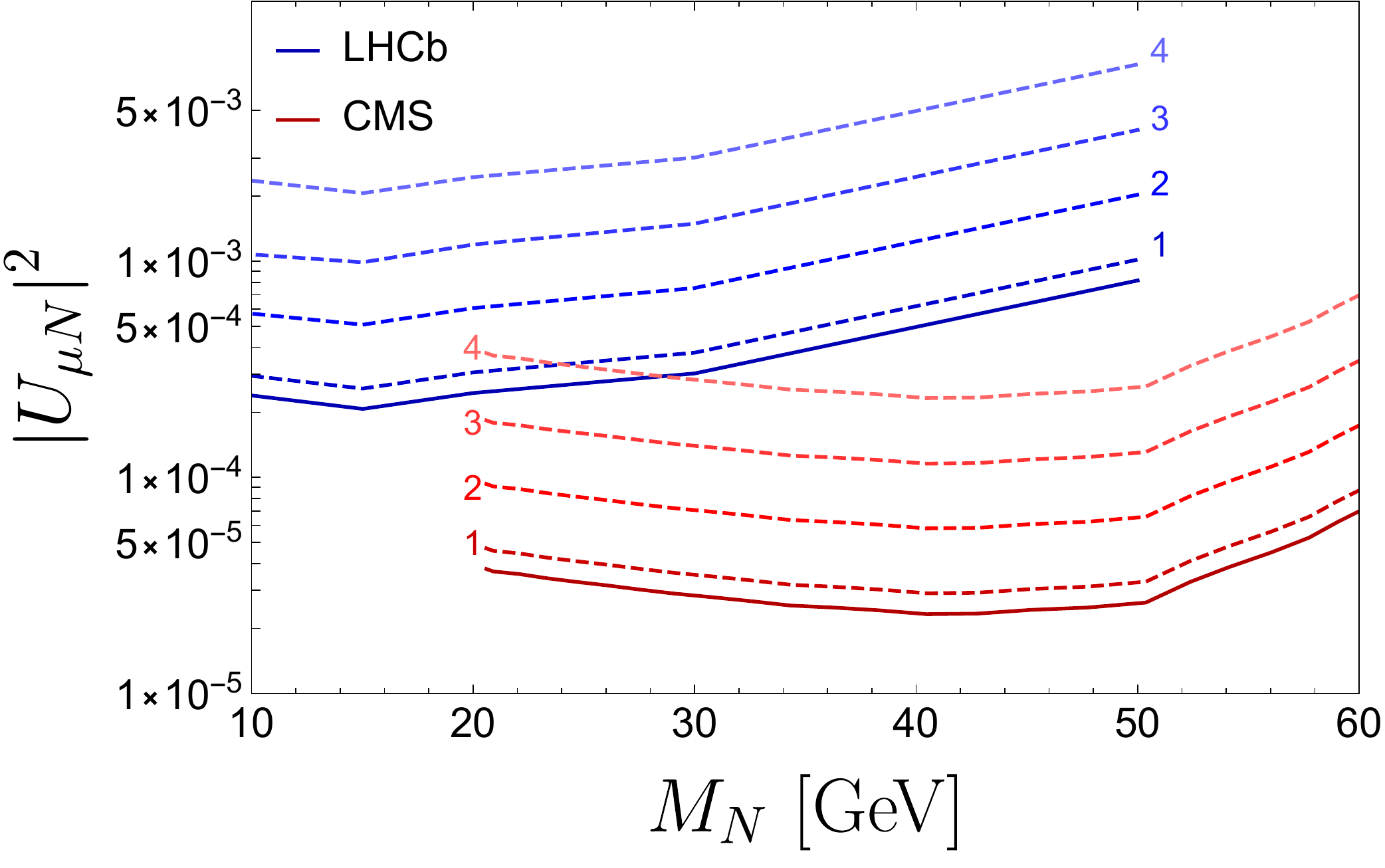} }}}
\caption[]{Rescaling of the bounds on $ |U_{\mu N}|^2$ from CMS\,\cite{CMS:2018jxx} (solid red line) and LHCb\,\cite{LHCb:2020wxx} (solid blue line), fixing the active$-$sterile mixings to the values corresponding to the green points $1-4$ in the ternary plot on the left. The orange squared point represents the single mixing case. }
\label{fig:recastDimuon}
\end{figure}

As a case of study for the use of Table~\ref{Tab:recastingMixingMajorana}, let us consider the CMS~\cite{CMS:2018jxx} and LHCb~\cite{LHCb:2020wxx} searches for the LNV dimuon channel $\mu^\pm \mu^\pm$.
For a given total active-sterile mixing $U^2$, 
we can display the full flavour mixing space in a ternary diagram, as in the left panel of Fig.~\ref{fig:recastDimuon}.
Then, the single mixing scenario constraints by both CMS and LHCb lie in the top corner. 
As we move along the ternary, we decrease the flavour strength to muons, so the bounds are relaxed, as shown in the right panel (dashed lines).
Here we chose just few benchmark points for the light HNL mass regime, although the same logic applies to the heavy one.

\begin{figure}[t!]
\begin{center}
\includegraphics[width=.49\textwidth]{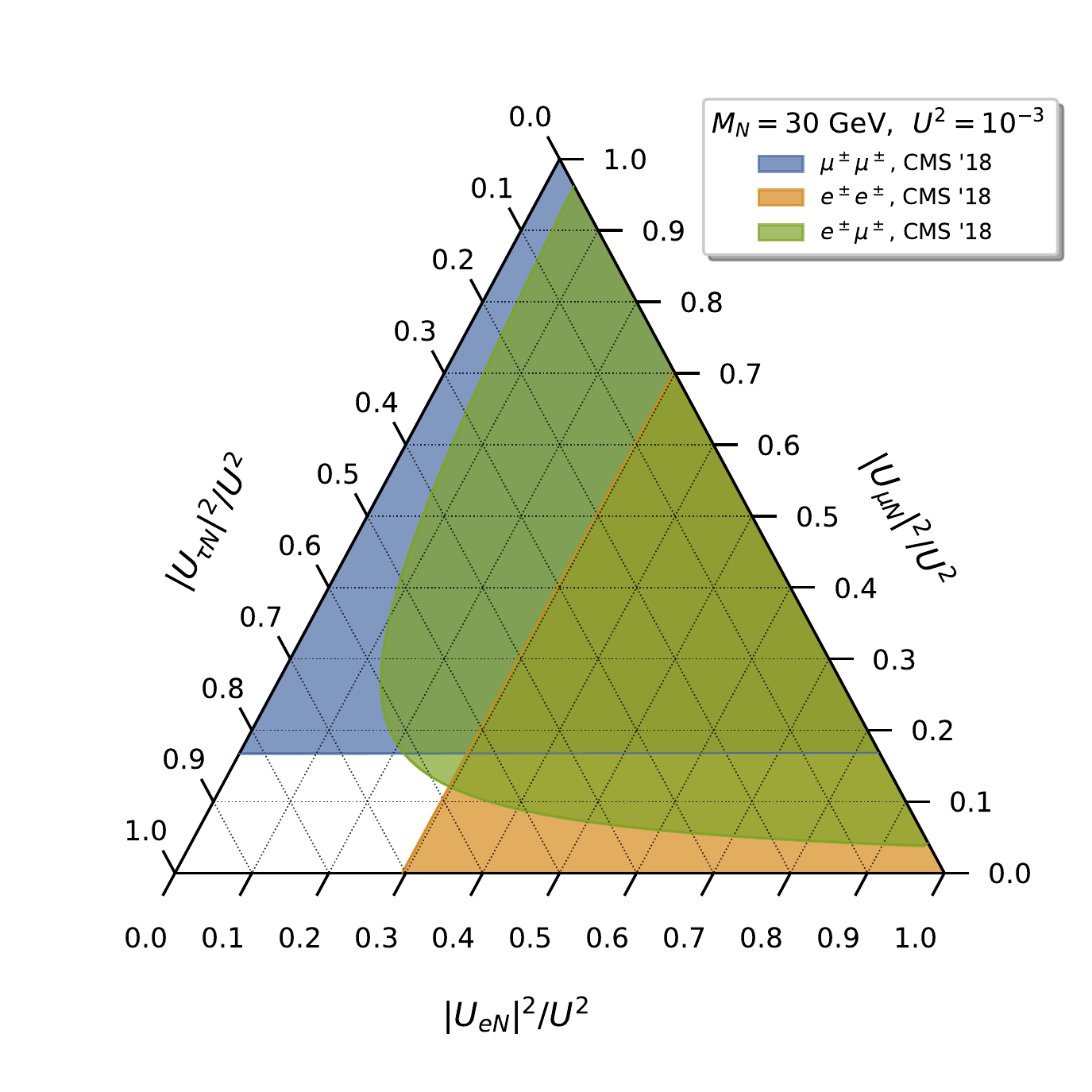}  \includegraphics[width=.49\textwidth]{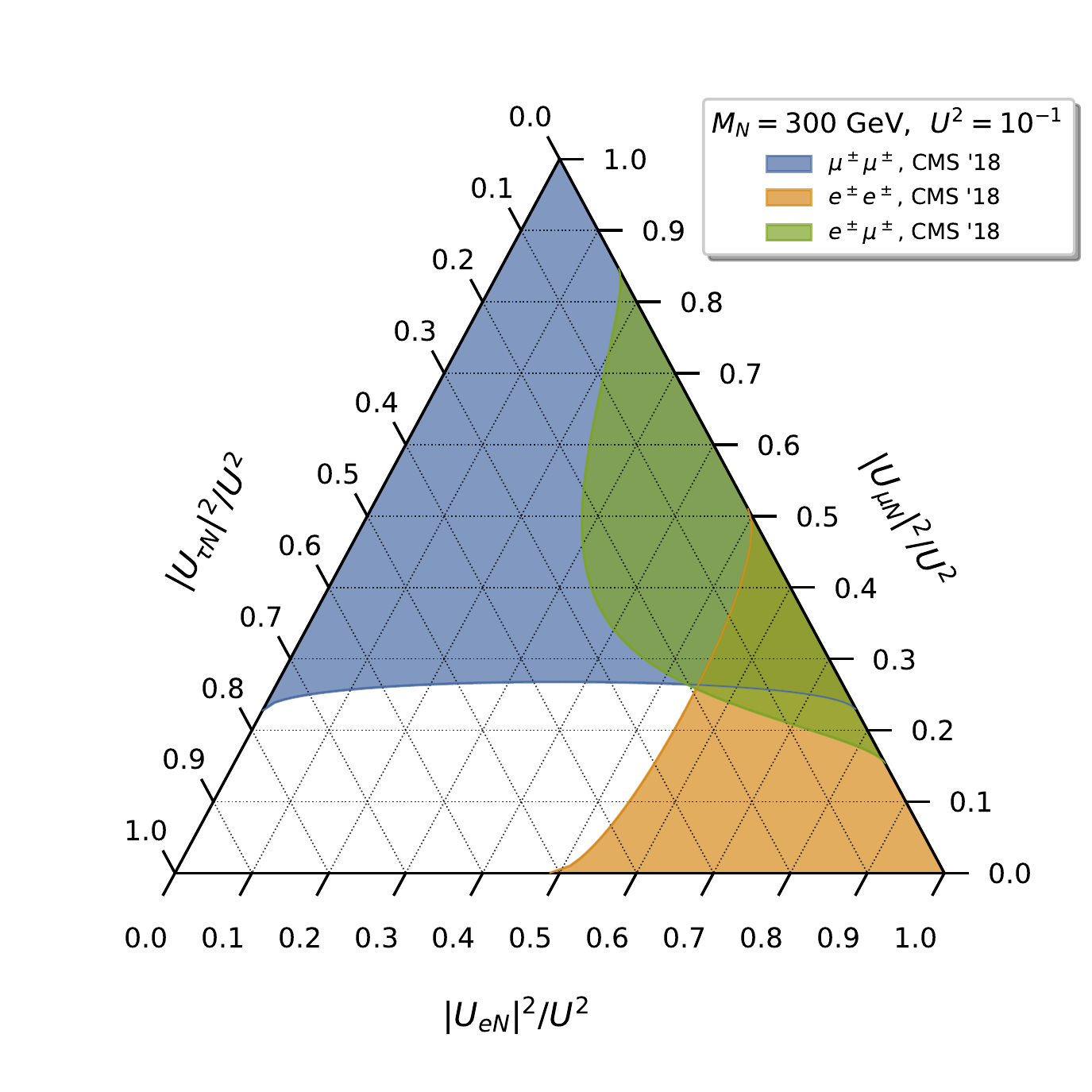}
\caption{
Combination of several bounds in a general mixing pattern for two benchmark points of mass and total mixing $U^2$.
Each bound correspond to the latest SS dilepton searches by CMS (Table~\ref{LHCexp}), which were derived within a single mixing scenario (or assuming $U_{\tau N}$ for the $e^\pm\mu^\pm$ channel).
The white area is still allowed by LHC searches.
}
\label{fig:ternaryFilled}
\end{center}
\end{figure}

As discussed above, the physical reason for the relaxation of these bounds is due to the new HNL decay channels in the generic mixing scenario. 
On the other hand, this also implies that the experimental searches for the other channels with different flavours might become relevant.
In order to show the interplay between different flavour channels, let us consider again a ternary diagram. 
We can understand it as a subspace of the parameter space with fixed values of $M_N$ and $U^2$, which is dissected in flavour space. 
Then, searches for dimuon channels will cover the area close to the $\varepsilon_\mu=1$ corner, becoming weaker as we move further away.
Equivalently, dielectron searches will cover the ternary from the $\varepsilon_e=1$ corner, while the $e^\pm\mu^\pm$ channel will cover the area in between.
This is depicted in Fig.~\ref{fig:ternaryFilled} for two benchmark points, one in the light regime ($M_N=30$~GeV) and one in the heavy regime ($M_N=300$~GeV).

Figure~\ref{fig:ternaryFilled} clearly shows the complementarity of the different dilepton searches, to which we could supplementary add the bounds from trilepton channels if the above mentioned concerns are solved.
If the combination of every channel covered all the area of the ternary, we could say that this $(M_N,U^2)$ 
point is excluded no matter which flavour mixing pattern we were considering.
This is not the case in any of the two examples in the figure, since the bottom left corner is still allowed by LHC.
Notice that this is always the case at present, since there are no LHC searches for HNLs mixing to the tau lepton, so the corner of $\varepsilon_\tau=1$ will always be allowed.
Although we already discussed that a single mixing scenario does not seem very natural for a realistic model including HNLs, closing this gap still motivates the need of performing dedicated searches in the tau sector.

On the other hand, it is important to stress that the benchmark points in Fig.~\ref{fig:ternaryFilled} are chosen just for illustrative purposes, since they are already excluded by LEP searches or global fit bounds.
Indeed, this is actually the case in a large part of the parameter space for the current LHC bounds discussed in Sec.~\ref{sec:status}.
Nevertheless, it is worth emphasizing that LHC sensitivities are expected to improve during the currently ongoing runs, pushing our knowledge about HNLs beyond present limits.
Therefore, combining the different LHC channels as we discussed here will become crucial in order to determine whether a heavy neutral lepton  with a given mass and mixing is completely excluded or not. 

\section{Beyond the single HNL}\label{sec:2HNL}

Should HNL exist in Nature, there is a priori no bound on their number and in general, BSM models involving HNLs do not introduce just one of them. 
For example, in the standard type-I seesaw model at least two HNL are needed to explain neutrino oscillation data. 
 In this work, in order to explore deviations from the single HNL hypothesis, we extend the framework considered in the previous section to include more neutral fermions focusing here on the minimal case of SM extension via two HNLs.  
In this case, the lepton mixing matrix is now a $5\times5$ unitary matrix

\begin{equation}
U_\nu = \left(\begin{array}{cc}
U_{\nu\nu}^{3\times3} & U_{\nu N}^{3\times2} \\
U_{N\nu}^{2\times3} & U_{NN}^{2\times2}
\end{array}\right)\,,
\end{equation}
with the fourth and fifth columns encoding  the mixings of both HNLs to the active leptons
\begin{equation}
U_{\nu N} = \left(\begin{array}{cc}
U_{e N_1} & U_{e N_2} \\
U_{\mu N_1} & U_{\mu N_2} \\
U_{\tau N_1} & U_{\tau N_2}
\end{array} \right)\,.
\end{equation}

For the sake of simplicity, we will assume in this section that the two HNLs mix to just a single flavour. 
In a sense, it can be seen as extending the interpretation of the bounds of Fig.~\ref{BoundsColliders} in horizontal in $U_{\nu N}$, to additional columns, while in Section~\ref{sec:flavour} we extended them in vertical, to additional flavours. 
The most general case with several HNLs and arbitrary mixing patterns can then be inferred as a combination of these two discussions.

In the case where only one of these HNLs is within  experimental reach or when there are several HNLs but with well-separated mass regimes, our conclusions derived following the single HNL scenario will apply to each of the HNLs.
Nevertheless, if two HNLs happen to be close in mass (as motivated by low-scale seesaw models~\cite{Gavela:2009cd,Ibarra:2010xw,Abada:2014vea}, resonant leptogenesis~\cite{Pilaftsis:2003gt} or ARS leptogenesis~\cite{Akhmedov:1998qx}), they could lead to interference effects and modify the results and the bounds obtained in the single HNL hypothesis.
Moreover, these modifications might affect both LNV and LNC branching ratios, and thus studying their correlation could shed light on the nature of the HNL (see for instance~\cite{Abada:2019bac} and references therein).

In this section we discuss how these effects could affect the LHC bounds obtained in the single HNL scenario from searches for LNV and LNC channels, and provide a recipe  to combine both results (on LNV and LNC searches) in order to bring forth  more robust bounds on HNLs parameter space. The recipe is  also  applicable to the case where HNLs interfere. 
We mostly focus on the LHCb~\cite{LHCb:2020wxx} results for the prompt dimuon channel, since this is the only available analysis addressing both SS and OS dilepton channels (cf.~Table~\ref{LHCexp}). 
More specifically, this search considers light HNLs which are dominantly produced from on-shell DY $W$ bosons in Fig.~\ref{DiagHNLatLHC}, that is,
\begin{equation}\label{Eq:Wprod}
p p \to W^+ \to \ell_\alpha^+ N_i \to \ell_\alpha^+ \ell_\beta^+ \, q \, \overline{q}^\prime\, , \quad p p \to W^+ \to \ell_\alpha^+ N_i \to \ell_\alpha^+ \ell_\beta^- \, q \, \overline{q}^\prime \, .
\end{equation}
In presence of just a single Majorana HNL, although the angular distributions will be different (see Appendix~\ref{app:DecayAmplitudes2HNL} for more details), the predictions for the total rates of these two processes are of similar size: equal for channels with $\alpha=\beta$ and twice as large for $\alpha\neq\beta$.
The reason for this difference is that the channel with crossed $\ell_\alpha$ and $\ell_\beta$ is also contributing  in the case of the LNV process. 
However, it must be added incoherently to the process since the rate is dominated by on-shell HNLs, fixing the momentum of the first lepton with the 2-body decay kinematics, and thus the interference becomes subdominant.
This means that for channels with $\alpha\neq\beta$, the total rate for the LNV process is enhanced by a factor of 2 with respect to the LNC one.
On the other hand, in channels with $\alpha=\beta$ there is an additional $1/2$ factor from having two identical particles, and thus we obtain the same total rate for both LNV and LNC.

When assuming the existence of two HNLs, we have two identical contributions to the total amplitude of the processes, one for each $N_i$.
The squared amplitude is then given as sum of the individual contributions of each HNL, plus a potential interference between $N_1$ and $N_2$ contributions. 
For each individual contribution to the amplitudes, we find that they are proportional to $U^*_{\alpha N_i} U^*_{\beta N_i}$ for the SS process and to $U^*_{\alpha N_i} U_{\beta N_i}^{\phantom *}$ for the OS one, proving convenient to define 
\begin{equation}
U_{\alpha N_i}=|U_{\alpha N_i}|\, e^{i \phi_{\alpha i}}\,,
\end{equation}
with $\phi_{\alpha i}\in [0,2\pi]$,  $\alpha = e,\mu,\tau$ and $i = 1,2$. 
In this way, the interference term resulting from both amplitudes will be proportional to
\begin{equation}
    \left|U_{\alpha N_1}\right|\left|U_{\alpha N_2}\right|\left|U_{\beta N_1}\right|\left|U_{\beta N_2}\right| e^{i \delta \phi^\pm} \, ,
\end{equation}
where we have defined $\delta \phi^\pm$ as follows:
\bea
\delta \phi^\pm= \left(\phi_{\alpha 2} - \phi_{\alpha 1}\right) \pm \left(\phi_{\beta 2} - \phi_{\beta 1}\right)\ ,
\eea 
with $+/-$ for the SS/OS channel.
Details with the complete analytical amplitude and decay rates, which we have additionally checked with {\it Whizard}~\cite{Brass:2019hvu}, can be found in Appendix~\ref{app:DecayAmplitudes2HNL}, both for the case with just one HNL and with two HNLs.

As  already stated, when the mass difference of the two HNLs is too large (compared to 
their decay width), 
the contribution of each HNL resonate independently and the interference is negligible.
In this case, the total rates for the LNV and LNC rates are related as in the case of the single HNL scenario.
Therefore, in the following we will assume the scenario when the interference effects can modify considerably the relative predictions between LNV and LNC rates, which corresponds to the case where both HNLs are close in mass. 
More specifically, we assume that the individual contribution of each HNL is of similar size,  $M_{N_1} \simeq M_{N_2} \equiv M_N, \Gamma_{N_1} \simeq \Gamma_{N_2} \equiv \Gamma_N$, 
 and also that $\left|U_{\alpha N_2}\right|\left|U_{\beta N_2}\right| = \left|U_{\alpha N_1}\right|\left|U_{\beta N_1}\right|$. However, we consider that the two HNL  mass splitting $\Delta M_N \equiv M_{N_2}-M_{N_1}$ could be different from zero as long as it is small compared to the decay width $\Gamma_N$. 
Notice that this kind of scenario appears naturally in low-scale seesaws due to the approximated lepton number conservation.
The only difference is that in these models the phases of the HNLs are fixed to be opposite, {\it i.e.}~$\delta\phi^+=\pi$ and $\delta\phi^-=0$ (the heavy neutrinos forming a pseudo-Dirac neutrino pair), and that $\Delta M_N$ is somehow related to light neutrino masses~\cite{Drewes:2019byd,Fernandez-Martinez:2022gsu}, while in the following we will let them free, in the spirit of the bottom-up approach described before.

Under these conditions, we can write the total decay rate driven by the two HNLs in the case of $W^+$ channel as
\begin{equation} \label{eq:decaywidthN1N2}
    \left.\Gamma\left(W^{+} \rightarrow \ell_{\alpha}^{+} \ell_{\beta}^{\pm} q \bar{q}^{\prime}\right)\right|_{N_{1} \& N_{2}}=2\,\mathcal{K}_+\left(y, \delta \phi^\pm \right)\,\left.\Gamma\left(W^{+} \rightarrow \ell_{\alpha}^{+} \ell_{\beta}^{\pm} q \bar{q}^{\prime}\right)\right|_{N_{1}}\  ,
\end{equation}
and equivalently for the $W^-$ channel with  $\mathcal{K}_-\left(y, \delta \phi^\pm \right)$ . 
Here, we have factorized the total rate in presence of only one HNL, and defined the modulation function
\begin{equation}
    \mathcal{K}_\pm\left(y, \delta \phi ^\pm\right) \equiv \left(1+\cos \delta \phi^\pm \frac{1}{1+y^{2}}\mp\sin \delta \phi^\pm \frac{y}{1+y^{2}}\right),
\end{equation}
with
\begin{equation}
    y  \simeq \frac{\Delta M_N}{\Gamma_N} \, .
\end{equation}
The function $ \mathcal{K}_\pm$ codifies the role of the interference. 
In the limit of $\Delta M_N\gg \Gamma_N$, the two HNL are too separated in mass, coherence is lost and the total contribution is just twice the single HNL contribution for both LNV and LNC. 
On the other hand, for $\Delta M_N < \Gamma_N$, the modulation function can take values from 0 (maximally destructive interference) to 2 (maximally constructive).
Thus, we are maximizing the effects of the interference between the two HNLs.
Moreover, these effects will be different for LNV and LNC, breaking the equal size prediction in the single HNL scenario.

\begin{figure}
\begin{center}
\includegraphics[width=.49\textwidth]{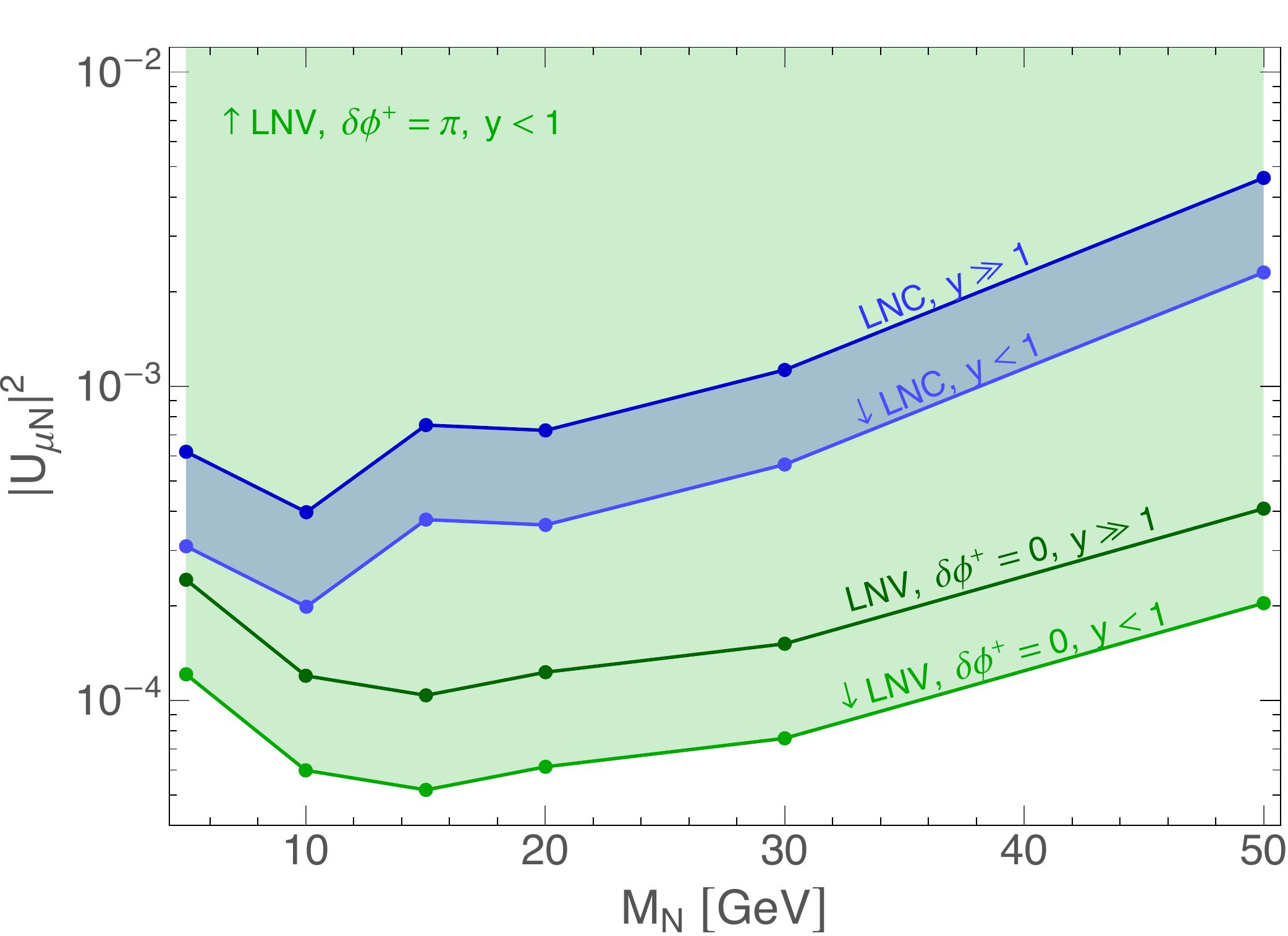}
\includegraphics[width=0.49\textwidth]{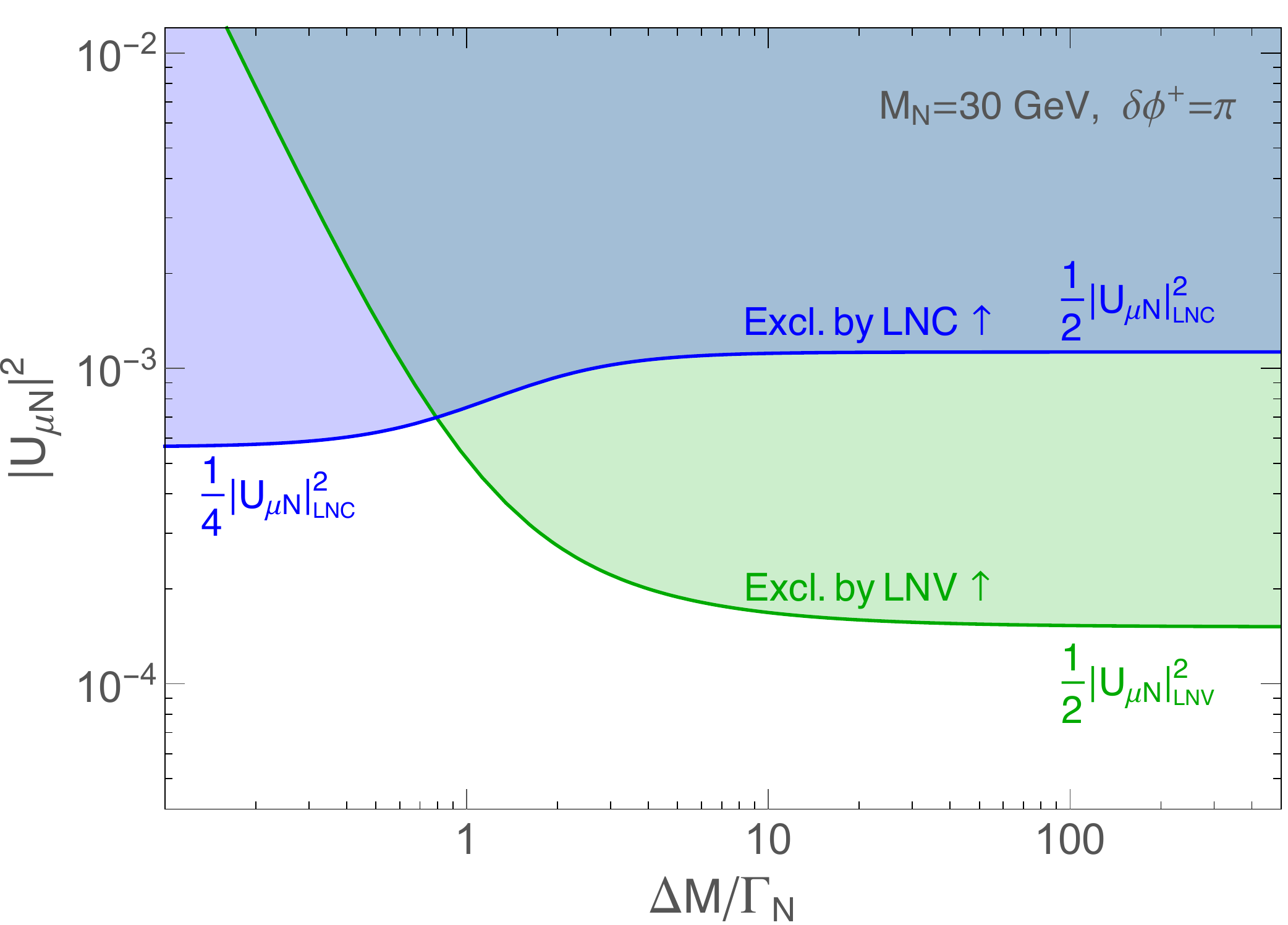}
\caption[]{Left: rescaling of the bounds on $ |U_{\mu N}|^2$ from LHCb\,\cite{LHCb:2020wxx} in the presence of two HNLs. Dark blue line is the LHCb bound in the LNC searches, while in lighter blue (lower curve) is the rescaled bound for $y= \Delta M_N/\Gamma_N<1$. The dark green line is the bound in the LNV searches, which can be relaxed (upper green region) if the $N_{1,2}$ form a pseudo$-$Dirac pair ($\delta \phi^+=\pi$, $y\ll1$), or strengthened (lower green region) if $\delta \phi^+=0$. Right: $ |U_{\mu N}|^2$ as a function of $y$, with $M_N=30$~GeV and $\delta \phi^+=\pi$. Blue (green) region is for LNC (LNV) channel with the thick lines corresponding to the (rescaled) LHCb bounds to the case of a low-scale seesaw with pseudo-Dirac HNL pair.}
\label{fig:LHCb2NHL}
\end{center}
\end{figure}

These modulation function can be used to simply recast the bounds derived by LHCb~\cite{LHCb:2020wxx}  under the assumption of a single HNL.
Noticing that LHCb searched only for the $W^+$ channel and given Eq.~\eqref{eq:sigmadilepton}, to recast these bounds to our  scenario with two HNLs, we need to rescale the mixing as
\begin{equation}
    \left| U_{\mu N} \right|^2 \rightarrow \left| U_{\mu N} \right|^2 \times 2 \mathcal{K}_+\left(y, \delta \phi^\pm \right) \, ,
\end{equation}
with $\delta \phi^- = 0$ in this channel with $\alpha=\beta$.
Following this modulation, we show in the left panel of Fig.~\ref{fig:LHCb2NHL} how the LHCb bounds might vary, depending on the  values of $y$ and the relative phases $\delta \phi^+$, for both LNV and LNC searches, which defines the green and blue bands, respectively.
The vertical axis needs to be understood now as the (squared) mixing of each of the HNLs to muons, which in absence of interference ($y\gg1$) is just a factor of two stronger with respect to the single HNL scenario.
For $y\ll1$, however, constructive interference can strengthen the bounds by up to an additional factor of two, while destructive interference could relax it, even avoid it completely in the case of LNV signals. 
The latter corresponds to the case where $\delta \phi^+=\pi$, which is precisely when the two HNLs have opposite phases (thus forming a pseudo-Dirac pair), as required by low-scale seesaws with approximated lepton number conservation.
Interestingly, the same choice of parameters that maximizes the destructive interference of LNV channels also maximizes the constructive interference for the LNC ones, making the bounds from the latter stronger.

This interplay prompts us to consider both channels at the same time, using their complementarity to set absolute bounds on the mixings that could not be avoided even with {\it ad-hoc} values of the parameters that maximize the interference. 
We show this in the right panel of Fig.~\ref{fig:LHCb2NHL} for a particular example of $M_N= 30$~GeV and opposite phases, $\delta \phi^+ = \pi$, as required by low-scale seesaws.
When the contributions of the two HNLs is maximally coherent ($y\ll1$), the LNV searches are avoided at the price of maximizing the LNC bounds.
If coherence is lost ($y\gg1$), then the stronger LNV bounds always dominate over the LNC ones. 
We see then that the largest possible mixing could be obtained in between the two cases, when the two tendencies cross over, which we can consider as an absolute bound on the mixing that cannot be avoided even with 2 interfering HNLs. 

To summarize, searches for LNC and LNV processes are important and complementary, since they can cover areas of the parameter space even in the case where  there is some interference (partially) cancelling any of the two channels.
This strongly motivates the need of searching for {\it both} LNC and LNV channels in parallel, even if the latter is more challenging experimentally, since combining both of them we could set more robust bounds on generic scenarios including more than one heavy neutral lepton.

\subsection*{CP violation}

When a quasi-degenerated pair of HNLs is considered, new CP-violating phases are introduced, which can induce differences in the decays of these particles to leptons over antileptons. 
If HNLs were discovered in processes such as those in Eq.~\eqref{Eq:Wprod}, and provided that enough events were collected, one way of measuring this potential CP asymmetry would be by defining the ratio~\cite{Najafi:2020dkp}
\begin{equation}\label{eq:ACP}
  A_{\mathrm{CP}}^\pm=\frac{{\rm BR}\left(W^- \rightarrow \ell_{\alpha}^{-} \ell_{\beta}^{\mp} \bar{q} q^{\prime} \right)-{\rm BR}\left(W^{+} \rightarrow \ell_{\alpha}^{+} \ell_{\beta}^{\pm} q \bar{q}^{\prime}\right)}{{\rm BR}\left(W^{-} \rightarrow \ell_{\alpha}^{-} \ell_{\beta}^{\mp} \bar{q} q^{\prime}\right)+{\rm BR}\left(W^{+} \rightarrow \ell_{\alpha}^{+} \ell_{\beta}^{\pm} q \bar{q}^{\prime}\right)} \, .
\end{equation}
Using our previous results for the $W$ decays, see Eq.~(\ref{eq:decaywidthN1N2}), it is straightforward to see that $A_{\rm CP}^\pm$ takes the simple form
\be \label{eq:ACP}
  A_{\rm CP}^\pm = \frac{y \, \sin \delta \phi^\pm }{1+ y^2 + \cos \delta \phi^\pm } \, ,
\ee
where $+/-$ denotes again LNV/LNC processes.

\begin{figure}[t!]
\centering
\includegraphics[width=0.6\textwidth]{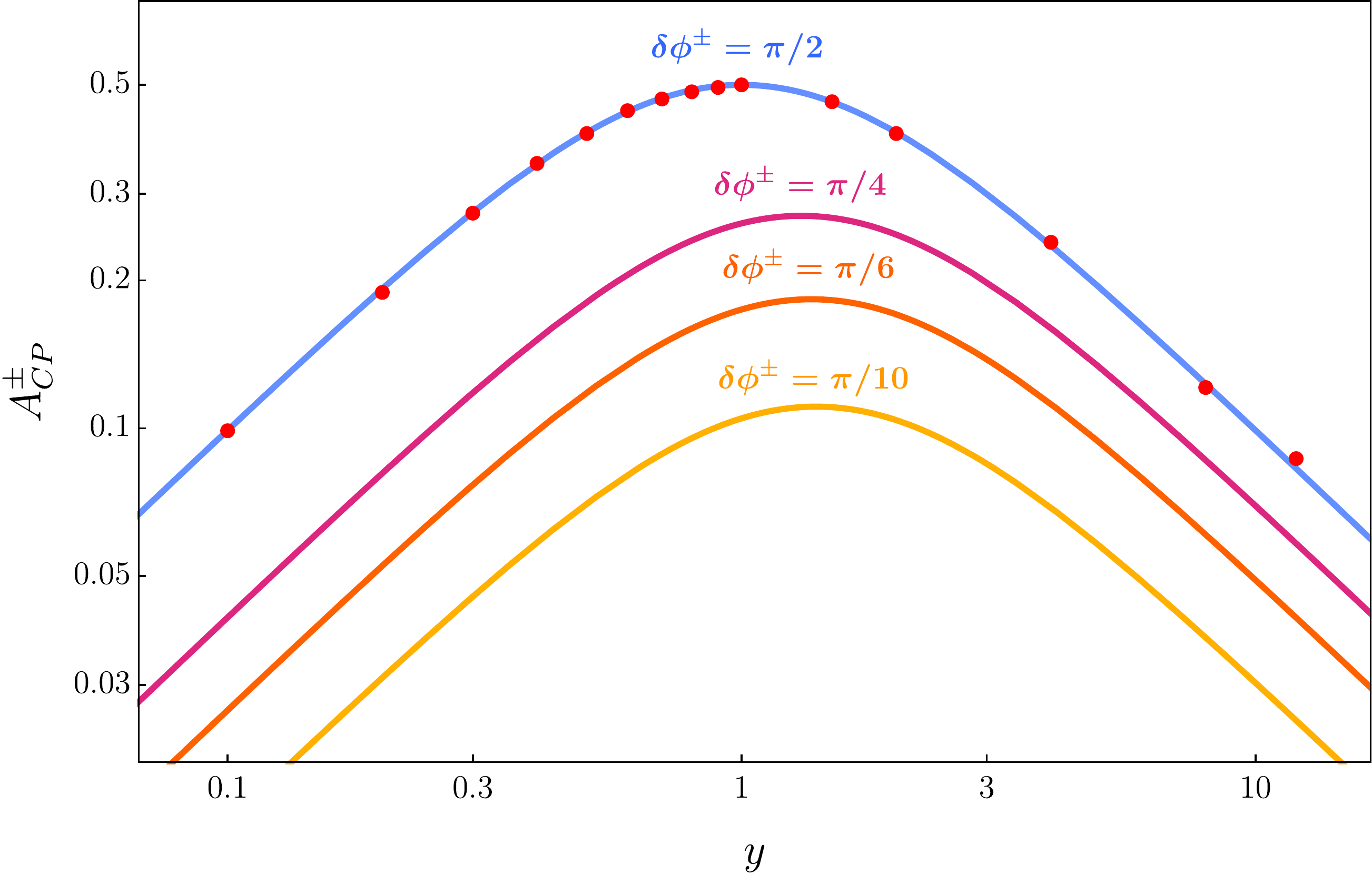}
\caption{
CP violating ratio $A_{\rm CP}^\pm$, defined in Eq.~(\ref{eq:ACP}), as function of the relevant ratio $y=\Delta M_N/\Gamma_N$ and for different choices of the relative phases $\delta\phi^\pm$.
Red dots were obtained with {\it Whizard} as a cross-check of our analytical results.}
\label{FigACP}
\end{figure}

As can be seen, this equation does not depend on the HNL masses, but on their mass difference $\Delta M_N$ (through $y= \Delta M_N/\Gamma_N$). 
It vanishes for the obvious case of $\delta \phi^\pm = 0$, since then there is no difference in both the $W$ decay and its CP equivalent; notice that it is also the case when  $y$ is close to zero or too large.
This is shown in Fig.~\ref{FigACP} where we display the CP asymmetry for different values of $\delta \phi^\pm$.

Notice that a similar computation was performed in Ref.~\cite{Najafi:2020dkp} with which our results agree but with some slight differences. 
On the one hand, we find a discrepancy when applying the narrow width approximation for the interference term (see App.~\ref{app:DecayAmplitudes2HNL} for more details).
On the other hand, we performed the computation considering very prompt heavy neutral leptons, so we did not take into account   their time evolution  and their possible oscillations before decaying.
Our approach is appropriate for heavy HNLs, while one should take into account the evolution effect for very light HNLs, thus with longer decay lifetimes, as it was done in Ref.~\cite{Najafi:2020dkp}.
Finally, we also point out that, in order to measure $A_{\rm CP}$ at a proton-proton collider, we must take into account that the production rates for $W^+$ and $W^-$ are not the same~\cite{Najafi:2020dkp}.

\section{Conclusions}\label{sec:concs}

In this work, we focused on LHC searches of heavy neutral leptons that decay promptly (short-lived). In most of the searches, which we summarized in Sec.~\ref{sec:status}, the obtained bounds were derived under the hypothesis of the existence of a single (usually Majorana) HNL that mixes with only one lepton flavour, while most of the BSM scenarios involving new neutral fermions require more than one HNL.
Moreover, unless some specific symmetries are present, the mixing pattern in these BSM scenarios is more complex, each HNL mixing in general with all charged leptons, and thus, the  bounds derived  from negative searches on the HNL parameter space have to be recast before being applied to a generic BSM scenario.

In this study we discussed how to recast the present experimentally obtained bounds on the parameter space, {\it i.e.}~active-sterile mixings $U_{\alpha N}$, $\alpha=e,\mu,\tau$, versus the  mass of the HNL, to the case of generic mixing to all active flavours as well as to the case with several HNLs.
The former was covered in Sec.~\ref{sec:flavour}, where we inspected the flavour dependencies of each of the channels searched for by the LHC, and stressed the importance of setting bounds not only on the mixings, but on the relevant combination of $|U_{\alpha N}|^2\times{\rm BR}$ (see Tables~\ref{Tab:recastingMixingMajorana} and \ref{Tab:recastingMixingDirac}).
Considering the bounds on this combination, we proposed a method to combine the results in flavour space, using the ternary diagrams in Fig.~\ref{fig:ternaryFilled} to conclude whether that area of parameter space is fully excluded, regardless of the assumed mixing pattern.

In the case with several heavy neutral leptons, we focused on the scenario when two HNLs are in the same mass regime, nearly degenerate or possibly forming a pseudo-Dirac neutrino pair, paying  special attention to the non-trivial role of the interference between their contributions.
To illustrate its importance, we focused on dilepton channels and moreover considered both channels with same and opposite charge of the final leptons, as it was done by LHCb~\cite{LHCb:2020wxx}.
We showed the complementary of the LNC and LNV searches and the importance of performing both of them in parallel. 
We stressed that by doing this, we are not only taking into account the two possible nature of a single HNL, Dirac or Majorana, but also covering the case when, for example, two Majorana HNLs exist and however they interfere destructively, suppressing the expected LNV signature.

To summarize, we have discussed the importance of going beyond simplified scenarios such as the single mixing hypothesis.
While they are useful for simplifying experimental analyses, they are not directly applicable to BSM models introducing HNLs.
Unfortunately, recasting the bounds of each experimental analysis to a given BSM scenario can be a tedious task.
In this work, we have proposed an alternative way of presenting the bounds on the parameter space of the HNLs, which under some approximations can be both directly constrained by experimental analyses and also easily recast to a generic BSM scenario.

\section*{Acknowledgements}

PE is grateful for the hospitality of the Pôle de Physique Théorique of the IJCLAB (Orsay) during the first stages of the project.
This project has received support from the European Union's Horizon 2020 research and innovation programme under the Marie Sk\l{}odowska-Curie grant agreement No.~860881 (HIDDe$\nu$ network).
The work of PE is supported by the Spanish grants PID2020-113775GB-I00 (AEI / 10.13039/501100011033), CIPROM/2021/054 (Generalitat Valenciana) and by the FPI grant PRE2018-084599.
XM also acknowledges partial support from the Spanish Research Agency (Agencia Estatal de Investigaci\'on) through the Grant IFT Centro de Excelencia Severo Ochoa No CEX2020-001007-S and Grant PID2019-108892RB-I00 funded by MCIN/AEI/10.13039/501100011033.

\appendix

\section{Decay process in presence of 2 HNL}
\label{app:DecayAmplitudes2HNL}

In this appendix we collect the relevant details for the computation of the $W$ boson decay into a HNL, followed by its semileptonic decay.
This is the relevant process for the searches performed by LHCb~\cite{LHCb:2020wxx} and that we discussed in Sec.~\ref{sec:2HNL} in the scenario with two heavy neutral leptons.

\subsection{Same sign leptons: $W^{+} \rightarrow \ell_{\alpha}^{+} \ell_{\beta}^{+} q' \bar{q}$}

We start with the decay rate for the process $W^{+} \rightarrow \ell_{\alpha}^{+} \ell_{\beta}^{+} q' \bar{q}$ mediated by two HNLs $N_{1,2}$ almost degenerate in mass, similar to that shown in Fig.~\ref{DiagHNLatLHC}, but without the initial quarks.
The amplitude reads
\begin{equation}
  \mathcal{M}^+=\sum_{i=1,2} \frac{g^{3}}{2 \sqrt{2}} U_{{\alpha} N_{i}}^{*} U_{{\beta} N_{i}}^{*} \frac{M_{N_i}}{M_{W}^{2}} \epsilon_{\mu}^{*} \frac{\overline{\ell_{\alpha}^{c}} \gamma^{\mu} \gamma^{\nu} P_{L} \ell_{\beta} \bar{q} \gamma_{\nu} P_{L} q^{\prime}}{p_{N}^{2}-M_{N_i}^{2}+i \Gamma_{N_i} M_{N_i}}\,,
\end{equation}
with $p_N = p_W − p_{\ell_\alpha}$. The channel with crossed $\ell_\alpha$ and $\ell_\beta$ gives rise to the same amplitude, but
it must be added incoherently to our process since the rate is dominated by on-shell $N_i$, and thus
the momentum of the first lepton is fixed by the 2-body decay kinematics. Therefore the two
processes do not interfere and we can neglect for the moment the crossed channel. The only
modification results in a factor of 2 in the rate.

Defining $U_{\alpha N_i}=|U_{\alpha N_i}|\, e^{i \phi_{\alpha i}}$, $\delta \phi^+= \left(\phi_{\alpha 2} - \phi_{\alpha 1}\right) + \left(\phi_{\beta 2} - \phi_{\beta 1}\right)$ and using the narrow width approximation (NWA), the squared matrix element becomes
  \begin{align}
  \hspace*{-.28cm}
    & \overline{|\mathcal{M}^+|}^{2}= \left[\frac{g^{3}}{2 \sqrt{2} M_{W}^{2}}\right]^{2} \pi\left(p_{\ell_{\beta}} \cdot p_{q}\right)\left(2 E_{\ell_{\alpha}} E_{q^{\prime}}+p_{\ell_{\alpha}} \cdot p_{q^{\prime}}\right) \times \nonumber \\
    &\Bigg\{ \left|U_{{\alpha} N_{1}}\right|^{2}\left|U_{{\beta} N_{1}}\right|^{2} \frac{M_{N_1}}{\Gamma_{N_1}} \delta\left(p_{N}^{2}-M_{N_1}^{2}\right)+\left|U_{{\alpha} N_{2}}\right|^{2}\left|U_{{\beta} N_{2}}\right|^{2} \frac{M_{N_2}}{\Gamma_{N_2}} \delta\left(p_{N}^{2}-M_{N_2}^{2}\right) \nonumber\\
    &+2\left|U_{{\alpha} N_{1}}\right|\left|U_{{\alpha} N_{2}}\right|\left|U_{{\beta} N_{1}}\right|\left|U_{{\beta} N_{2}}\right| M_{N_1} M_{N_2}\left[\delta\left(p_{N}^{2}-M_{N_1}^{2}\right)+\delta\left(p_{N}^{2}-M_{N_2}^{2}\right)\right]\nonumber \\
    & \left. \left[\cos \delta \phi^+\,\frac{\Gamma_{N_1} M_{N_1}+\Gamma_{N_2} M_{N_2}}{\left(\Delta M_N^{2}\right)^{2}+\left(\Gamma_{N_1} M_{N_1}+\Gamma_{N_2} M_{N_2}\right)^{2}}
    -\sin \delta \phi^+\,\frac{\Delta M_N^{2}}{\left(\Delta M_N^{2}\right)^{2}+\left(\Gamma_{N_1} M_{N_1}+\Gamma_{N_2} M_{N_2}\right)^{2}}\right] \right\} \, ,
  \end{align}
where $ \Delta M_N^{2} \equiv M_{N_2}^{2}-M_{N_1}^{2}$.
Notice that, for the interference term, we used the NWA as follows:
\begin{align} \label{eq:interffrac}
        & \frac{1}{\left(p_{N}^{2}-M_{N_1}^{2}+i \Gamma_{N_1} M_{N_1}\right)\left(p_{N}^{2}-M_{N_2}^{2}- i \Gamma_{N_2} M_{N_2}\right)} = \nonumber \\
        & \hspace{4.45cm} \frac{\pi\left(\Gamma_{N_2} M_{N_2}+\Gamma_{N_1} M_{N_1}\right)}{\left(\Delta M_N^{2}\right)^{2}+\left(\Gamma_{N_1} M_{N_1}+\Gamma_{N_2} M_{N_2}\right)^{2}}\left[\delta\left(p_{N}^{2}-M_{N_1}^{2}\right)+\delta\left(p_{N}^{2}-M_{N_2}^{2}\right)\right] \nonumber\\
        & \hspace{4.cm} +\frac{i \pi \Delta M_N^{2}}{\left(\Delta M_{N}^2\right)^{2}+\left(\Gamma_{N_1} M_{N_1}+\Gamma_{N_2} M_{N_2}\right)^{2}}\left[\delta\left(p_{N}^{2}-M_{N_1}^{2}\right)+\delta\left(p_{N}^{2}-M_{N_2}^{2}\right)\right],
\end{align}
which differs from the expression in Ref.~\cite{Najafi:2020dkp}, as discussed in Section \ref{sec:2HNL}. 
Assuming $M_{N_1} \simeq M_{N_2} \equiv M_N, \Gamma_{N_1} \simeq \Gamma_{N_2} \equiv \Gamma_N$ and $\Delta M_N \equiv M_{N_2}-M_{N_1} \neq 0$, and considering that $\abs{U_{\alpha N_1}} \abs{U_{\beta  N_1}} = \abs{U_{\alpha N_2}} \abs{U_{\beta N_2}} \equiv \abs{U_{\alpha N}} \abs{U_{\beta N}}$, we get
\begin{align}\label{eq:MplusNWA}
        \overline{|\mathcal{M}^+|}^{2} \simeq &\left[\frac{g^{3}}{2 \sqrt{2} M_{W}^{2}}\right]^{2} \pi\left(p_{\ell_{\beta}} \cdot p_{q}\right)\left(2 E_{{\alpha}} E_{q^{\prime}}+p_{{\alpha}} \cdot p_{q^{\prime}}\right) \delta\left(p_{N}^{2}-M_N^{2}\right) \frac{M_N}{\Gamma_N}\left|U_{{\alpha} N}\right|^{2}\left|U_{{\beta} N}\right|^{2} \nonumber \\
        & \times 2 \Biggl\{ 1 + 2 \left[2 \cos \delta \phi^+ \frac{M_N^{2} \Gamma_N^{2}}{\left(\Delta M_N^{2}\right)^{2}+4 \Gamma_N^{2} M_N^{2}}  - \sin \delta \phi^+ \frac{M_N \Gamma_N \Delta M_N^{2}}{\left(\Delta M_N^{2}\right)^{2}+4 \Gamma_N^{2} M_N^{2}}\right] \Biggr\} \, .
\end{align}
We observe that the squared amplitude in the case of only one sterile neutrino factorizes out. 
Integrating over the phase space and after factorizing the decay width in the single HNL framework, it is straightforward to obtain
\begin{equation} \label{eq:decaywidthN1N2app}
    \Gamma\left(W^{+} \rightarrow \ell_{\alpha}^{+} \ell_{\beta}^{+} q' \bar{q}\right)=2\left(1+\cos \delta \phi^+ \frac{1}{1+y^{2}}-\sin \delta \phi^+ \frac{y}{1+y^{2}}\right) 
    \left.\Gamma\left(W^{+} \rightarrow \ell_{\alpha}^{+} \ell_{\beta}^{+} q' \bar{q}\right)\right|_{N_{1}}  \, ,
\end{equation}
where we have defined
\begin{equation} \label{eq:defy}
    y \equiv \frac{\Delta M_N^{2}}{2 M_N \Gamma_N} \simeq \frac{M_{N_2}^{2}-M_{N_1}^{2}}{\left(M_{N_1}+M_{N_2}\right) \Gamma_N}=\frac{\Delta M_N}{\Gamma_N} \, .
\end{equation}
Notice that we obtain the same result of Ref.~\cite{Das:2017hmg}. This is because we are using the NWA, which corresponds to consider on-shell HNL, leading to the same result after integrating over $t$, the time  evolution of the intermediate HNL, from $0$ to $\infty$. In fact the factor $1/\Gamma$ coming from the NWA is equivalent to the factor $\int^\infty_0  \abs{e^{-\Gamma t /2}}^2 =1/\Gamma$, and the interference part of Eq.~\eqref{eq:MplusNWA} coincides with the finding  of Ref.~\cite{Das:2017hmg}.

\subsection{Different sign leptons: $W^+ \to \ell_\alpha^+ \ell_\beta^- q' \bar{q}$}

In this appendix, the decay rate for the process $W^+ \to \ell_\alpha^+ \ell_\beta^- q' \bar{q}$ mediated by the two HNLs $N_{1,2}$ almost degenerate in mass is computed. 
The amplitude of this process is given by
\be \label{eq:amplitude}
\mathcal{M}^-= \sum_{i=1,2}\frac{g^3}{2\sqrt{2} M_W^2} U_{\alpha N_i}^* U_{\beta N_i} \epsilon_\mu^* \frac{\bar{\ell}_\beta \gmuu \slashed{p}_N \gnuu P_L \ell_\alpha^+ \bar q \gnud P_L q' }{p_N^2 - M_{N_i}^2+ i \Gamma_{N_i} M_{N_i}} \, ,
\ee
where, again, $p_N = p_W - p_{\ell_\alpha}$.
With the help of {\it FeynCalc}~\cite{Shtabovenko:2016sxi}, we obtain for the squared amplitude, 
  \begin{align}
    \left| \overline{\mathcal{M}^-} \right|^2 = & \left( \frac{g^3}{2 \sqrt{2} M_W^2} \right)^2 \left\{ \abs{U_{\alpha N_1}}^2 \abs{U_{\beta N_1}}^2 \frac{K_{11}}{\left(p_N^2 - M_{N_1}^2 \right)^2 + \Gamma_{N_1}^2 \, M_{N_1}^2} \right. \nonumber\\
     + &\abs{U_{\alpha N_2}}^2 \abs{U_{\beta N_2}}^2 \frac{K_{22}}{\left(p_N^2 - M_{N_2}^2\right)^2 + \Gamma_{N_2}^2 \, M_{N_2}^2}  \nonumber\\
     + &2 \left. \text{Re} \left[ U_{\alpha N_1}^* U_{\beta N_1} U_{\alpha N_2} U_{\beta N_2}^* \frac{K_{12}}{\left(p_N^2 - M_{N_1}^2 + i \, \Gamma_{N_1} \, M_{N_1}\right) \left(p_N^2 - M_{N_2}^2 - i \, \Gamma_{N_2} \, M_{N_2}\right)} \right] \right\} \, ,
  \end{align}
with the $K_{ij}$ factors defined as
\begin{align}
    K_{ij} = & - \frac{16}{M_W} \left( p_{\beta} \cdot p_{q^\prime} \right) \biggl\{ M_W \, \left( p_{N_i} \cdot p_{N_j} \right) \Bigl[ \left( p_{\alpha} \cdot p_{q} \right) + 2 \, E_{\alpha} \, E_q \Bigr]  \biggr. - 2 \, E_{\alpha} \left[ \left(p_{N_i} \cdot p_W \right) \left(p_{N_j} \cdot p_q \right) \right. \nonumber \\
    & \left. + \left(p_{N_j} \cdot p_W \right) \left(p_{N_i} \cdot p_q \right) \right] \biggl. - M_W \left[ \left(p_{N_i} \cdot p_q \right) \left(p_{N_j} \cdot p_{\alpha} \right) + \left(p_{N_j} \cdot p_q \right) \left(p_{N_i} \cdot p_{\alpha} \right) \right] \biggr\}\,.
\end{align}
Assuming as before that $M_{N_1} \approx M_{N_2} \equiv M_N$ and $\Gamma_{N_1} \approx \Gamma_{N_2} \equiv \Gamma_N$, then $K_{11} = K_{22} = K_{12} \equiv K$. Also, considering that $\abs{U_{\alpha N_1}} \abs{U_{\beta  N_1}} = \abs{U_{\alpha N_2}} \abs{U_{\beta N_2}} \equiv \abs{U_{\alpha N}} \abs{U_{\beta N}}$, we can simplify the expression notably
\begin{align}
  \left| \overline{\mathcal{M}^-} \right|^2 = &  \left( \frac{g^3}{2 \sqrt{2} M_W^2} \right)^2 \abs{U_{\alpha N}}^2 \abs{U_{\beta N}}^2 \, 2 \, \left\{ \frac{K}{\left(p_N^2 - M_N^2 \right)^2 + \Gamma_N^2 \, M_N^2} \right.\nonumber \\    %
    +& \left. \text{Re} \left[ \frac{K \, e^{i \delta \phi^{-}}}{\left(p_N^2 - M_{N_1}^2 + i \, \Gamma_{N_1} \, M_{N_1}\right) \left(p_N^2 - M_{N_2}^2 - i \, \Gamma_{N_2} \, M_{N_2}\right)} \right] \right\} \, ,
\end{align}
with $\delta \phi^-= \left(\phi_{\alpha 2} - \phi_{\alpha 1}\right) - \left(\phi_{\beta 2} - \phi_{\beta 1}\right)$. And using the narrow width approximation, we get 
  \begin{align}
\left| \overline{\mathcal{M}^-} \right|^2 = & \,  2 \left( \frac{g^3}{2 \sqrt{2} M_W^2} \right)^2 \abs{U_{\alpha N}}^2 \abs{U_{\beta N}}^2 \, K \, \frac{\pi}{\Gamma_N \, M_N} \delta \left( p_N^2 - M_N^2 \right) \nonumber\\
    & \left\{ 1 +  \text{Re} \biggl[ 2 \, \Gamma_N \, M_N \frac{ 2 \, \Gamma_N \, M_N + i \, \Delta M_N^2 }{\left( \Delta M_N^2 \right)^2 + 4 \, \Gamma_N^2 \, M_N^2} e^{i \, \delta \phi^{-}}  \biggr] \right\} \, ,
  \end{align}
Finally, it is straightforward to obtain the total decay width in terms of the decay width mediated by just one HNL,
\be
    \Gamma \left( W^+ \to \ell_\alpha^+ \ell_\beta^- q' \bar{q} \right) = 2 \( 1 + \cos{\delta \phi^{-}} \frac{1}{1+y^2} - \sin{\delta \phi^{-}}\frac{y}{1+y^2}\) \times \Gamma \left( W^+ \to \ell_\alpha^+ \ell_\beta^- q' \bar{q} \right)\Bigl. \Bigr|_{N_1} \, ,
\ee
with $y$ defined in Eq.~\eqref{eq:defy}.

\bibliographystyle{utphys}
\bibliography{Refs}

\providecommand{\href}[2]{#2}\begingroup\raggedright\begin{thebibliography}{100}

\bibitem{Minkowski:1977sc}
P.~Minkowski, \emph{{$\mu \to e\gamma$ at a Rate of One Out of $10^{9}$ Muon
  Decays?}}, \href{https://doi.org/10.1016/0370-2693(77)90435-X}{\emph{Phys.
  Lett. B} {\bfseries 67} (1977) 421--428}.

\bibitem{Yanagida:1979as}
T.~Yanagida, \emph{{Horizontal gauge symmetry and masses of neutrinos}},
  {\emph{Conf. Proc. C} {\bfseries 7902131} (1979) 95--99}.

\bibitem{Glashow:1979nm}
S.~L. Glashow, \emph{{The Future of Elementary Particle Physics}},
  \href{https://doi.org/10.1007/978-1-4684-7197-7_15}{\emph{NATO Sci. Ser. B}
  {\bfseries 61} (1980) 687}.

\bibitem{Gell-Mann:1979vob}
M.~Gell-Mann, P.~Ramond and R.~Slansky, \emph{{Complex Spinors and Unified
  Theories}}, {\emph{Conf. Proc. C} {\bfseries 790927} (1979) 315--321},
  [\href{https://arxiv.org/abs/1306.4669}{{\ttfamily 1306.4669}}].

\bibitem{Mohapatra:1979ia}
R.~N. Mohapatra and G.~Senjanovic, \emph{{Neutrino Mass and Spontaneous Parity
  Nonconservation}},
  \href{https://doi.org/10.1103/PhysRevLett.44.912}{\emph{Phys. Rev. Lett.}
  {\bfseries 44} (1980) 912}.

\bibitem{Schechter:1980gr}
J.~Schechter and J.~W.~F. Valle, \emph{{Neutrino Masses in $SU(2) \times U(1)$
  Theories}}, \href{https://doi.org/10.1103/PhysRevD.22.2227}{\emph{Phys. Rev.
  D} {\bfseries 22} (1980) 2227}.

\bibitem{Gronau:1984ct}
M.~Gronau, C.~N. Leung and J.~L. Rosner, \emph{{Extending Limits on Neutral
  Heavy Leptons}}, \href{https://doi.org/10.1103/PhysRevD.29.2539}{\emph{Phys.
  Rev. D} {\bfseries 29} (1984) 2539}.

\bibitem{Mohapatra:1986bd}
R.~N. Mohapatra and J.~W.~F. Valle, \emph{{Neutrino Mass and Baryon Number
  Nonconservation in Superstring Models}},
  \href{https://doi.org/10.1103/PhysRevD.34.1642}{\emph{Phys. Rev. D}
  {\bfseries 34} (1986) 1642}.

\bibitem{Barr:2003nn}
S.~M. Barr, \emph{{A Different seesaw formula for neutrino masses}},
  \href{https://doi.org/10.1103/PhysRevLett.92.101601}{\emph{Phys. Rev. Lett.}
  {\bfseries 92} (2004) 101601},
  [\href{https://arxiv.org/abs/hep-ph/0309152}{{\ttfamily hep-ph/0309152}}].

\bibitem{Malinsky:2005bi}
M.~Malinsky, J.~C. Romao and J.~W.~F. Valle, \emph{{Novel supersymmetric
  $SO(10)$ seesaw mechanism}},
  \href{https://doi.org/10.1103/PhysRevLett.95.161801}{\emph{Phys. Rev. Lett.}
  {\bfseries 95} (2005) 161801},
  [\href{https://arxiv.org/abs/hep-ph/0506296}{{\ttfamily hep-ph/0506296}}].

\bibitem{Pontecorvo:1957cp}
B.~Pontecorvo, \emph{{Mesonium and anti-mesonium}}, {\emph{Sov. Phys. JETP}
  {\bfseries 6} (1957) 429}.

\bibitem{Maki:1962mu}
Z.~Maki, M.~Nakagawa and S.~Sakata, \emph{{Remarks on the unified model of
  elementary particles}}, \href{https://doi.org/10.1143/PTP.28.870}{\emph{Prog.
  Theor. Phys.} {\bfseries 28} (1962) 870--880}.

\bibitem{Atre:2009rg}
A.~Atre, T.~Han, S.~Pascoli and B.~Zhang, \emph{{The Search for Heavy Majorana
  Neutrinos}}, \href{https://doi.org/10.1088/1126-6708/2009/05/030}{\emph{JHEP}
  {\bfseries 05} (2009) 030},
  [\href{https://arxiv.org/abs/0901.3589}{{\ttfamily 0901.3589}}].

\bibitem{Abada:2017jjx}
A.~Abada, V.~De~Romeri, M.~Lucente, A.~M. Teixeira and T.~Toma,
  \emph{{Effective Majorana mass matrix from tau and pseudoscalar meson lepton
  number violating decays}},
  \href{https://doi.org/10.1007/JHEP02(2018)169}{\emph{JHEP} {\bfseries 02}
  (2018) 169}, [\href{https://arxiv.org/abs/1712.03984}{{\ttfamily
  1712.03984}}].

\bibitem{Bolton:2019pcu}
P.~D. Bolton, F.~F. Deppisch and P.~S. Bhupal~Dev, \emph{{Neutrinoless double
  beta decay versus other probes of heavy sterile neutrinos}},
  \href{https://doi.org/10.1007/JHEP03(2020)170}{\emph{JHEP} {\bfseries 03}
  (2020) 170}, [\href{https://arxiv.org/abs/1912.03058}{{\ttfamily
  1912.03058}}].

\bibitem{Agrawal:2021dbo}
P.~Agrawal et~al., \emph{{Feebly-interacting particles: FIPs 2020 workshop
  report}}, \href{https://doi.org/10.1140/epjc/s10052-021-09703-7}{\emph{Eur.
  Phys. J. C} {\bfseries 81} (2021) 1015},
  [\href{https://arxiv.org/abs/2102.12143}{{\ttfamily 2102.12143}}].

\bibitem{Pilaftsis:1991ug}
A.~Pilaftsis, \emph{{Radiatively induced neutrino masses and large Higgs
  neutrino couplings in the standard model with Majorana fields}},
  \href{https://doi.org/10.1007/BF01482590}{\emph{Z. Phys. C} {\bfseries 55}
  (1992) 275--282}, [\href{https://arxiv.org/abs/hep-ph/9901206}{{\ttfamily
  hep-ph/9901206}}].

\bibitem{delAguila:2007qnc}
F.~del {\'A}guila, J.~A. Aguilar-Saavedra and R.~Pittau, \emph{{Heavy neutrino
  signals at large hadron colliders}},
  \href{https://doi.org/10.1088/1126-6708/2007/10/047}{\emph{JHEP} {\bfseries
  10} (2007) 047}, [\href{https://arxiv.org/abs/hep-ph/0703261}{{\ttfamily
  hep-ph/0703261}}].

\bibitem{delAguila:2008cj}
F.~del {\'A}guila and J.~A. Aguilar-Saavedra, \emph{{Distinguishing seesaw
  models at LHC with multi-lepton signals}},
  \href{https://doi.org/10.1016/j.nuclphysb.2008.12.029}{\emph{Nucl. Phys.}
  {\bfseries B813} (2009) 22--90},
  [\href{https://arxiv.org/abs/0808.2468}{{\ttfamily 0808.2468}}].

\bibitem{delAguila:2008hw}
F.~del {\'A}guila and J.~A. Aguilar-Saavedra, \emph{{Electroweak scale seesaw
  and heavy Dirac neutrino signals at LHC}},
  \href{https://doi.org/10.1016/j.physletb.2009.01.010}{\emph{Phys. Lett.}
  {\bfseries B672} (2009) 158--165},
  [\href{https://arxiv.org/abs/0809.2096}{{\ttfamily 0809.2096}}].

\bibitem{Nemevsek:2011hz}
M.~Nemev{\v s}ek, F.~Nesti, G.~Senjanovi{\'c} and Y.~Zhang, \emph{{First Limits
  on Left-Right Symmetry Scale from LHC Data}},
  \href{https://doi.org/10.1103/PhysRevD.83.115014}{\emph{Phys. Rev.}
  {\bfseries D83} (2011) 115014},
  [\href{https://arxiv.org/abs/1103.1627}{{\ttfamily 1103.1627}}].

\bibitem{BhupalDev:2012zg}
P.~S. Bhupal~Dev, R.~Franceschini and R.~N. Mohapatra, \emph{{Bounds on TeV
  Seesaw Models from LHC Higgs Data}},
  \href{https://doi.org/10.1103/PhysRevD.86.093010}{\emph{Phys. Rev. D}
  {\bfseries 86} (2012) 093010},
  [\href{https://arxiv.org/abs/1207.2756}{{\ttfamily 1207.2756}}].

\bibitem{Cely:2012bz}
C.~G. Cely, A.~Ibarra, E.~Molinaro and S.~T. Petcov, \emph{{Higgs Decays in the
  Low Scale Type I See-Saw Model}},
  \href{https://doi.org/10.1016/j.physletb.2012.11.026}{\emph{Phys. Lett. B}
  {\bfseries 718} (2013) 957--964},
  [\href{https://arxiv.org/abs/1208.3654}{{\ttfamily 1208.3654}}].

\bibitem{Helo:2013esa}
J.~C. Helo, M.~Hirsch and S.~Kovalenko, \emph{{Heavy neutrino searches at the
  LHC with displaced vertices}},
  \href{https://doi.org/10.1103/PhysRevD.89.073005,
  10.1103/PhysRevD.93.099902}{\emph{Phys. Rev.} {\bfseries D89} (2014) 073005},
  [\href{https://arxiv.org/abs/1312.2900}{{\ttfamily 1312.2900}}].

\bibitem{Dev:2013wba}
P.~S.~B. Dev, A.~Pilaftsis and U.-k. Yang, \emph{{New Production Mechanism for
  Heavy Neutrinos at the LHC}},
  \href{https://doi.org/10.1103/PhysRevLett.112.081801}{\emph{Phys. Rev. Lett.}
  {\bfseries 112} (2014) 081801},
  [\href{https://arxiv.org/abs/1308.2209}{{\ttfamily 1308.2209}}].

\bibitem{Das:2014jxa}
A.~Das, P.~S. Bhupal~Dev and N.~Okada, \emph{{Direct bounds on electroweak
  scale pseudo-Dirac neutrinos from $\sqrt s=8$ TeV LHC data}},
  \href{https://doi.org/10.1016/j.physletb.2014.06.058}{\emph{Phys. Lett.}
  {\bfseries B735} (2014) 364--370},
  [\href{https://arxiv.org/abs/1405.0177}{{\ttfamily 1405.0177}}].

\bibitem{Alva:2014gxa}
D.~Alva, T.~Han and R.~Ruiz, \emph{{Heavy Majorana neutrinos from $W\gamma$
  fusion at hadron colliders}},
  \href{https://doi.org/10.1007/JHEP02(2015)072}{\emph{JHEP} {\bfseries 02}
  (2015) 072}, [\href{https://arxiv.org/abs/1411.7305}{{\ttfamily 1411.7305}}].

\bibitem{Izaguirre:2015pga}
E.~Izaguirre and B.~Shuve, \emph{{Multilepton and Lepton Jet Probes of
  Sub-Weak-Scale Right-Handed Neutrinos}},
  \href{https://doi.org/10.1103/PhysRevD.91.093010}{\emph{Phys. Rev.}
  {\bfseries D91} (2015) 093010},
  [\href{https://arxiv.org/abs/1504.02470}{{\ttfamily 1504.02470}}].

\bibitem{Deppisch:2015qwa}
F.~F. Deppisch, P.~S. Bhupal~Dev and A.~Pilaftsis, \emph{{Neutrinos and
  Collider Physics}},
  \href{https://doi.org/10.1088/1367-2630/17/7/075019}{\emph{New J. Phys.}
  {\bfseries 17} (2015) 075019},
  [\href{https://arxiv.org/abs/1502.06541}{{\ttfamily 1502.06541}}].

\bibitem{Maiezza:2015lza}
A.~Maiezza, M.~Nemev{\v s}ek and F.~Nesti, \emph{{Lepton Number Violation in
  Higgs Decay at LHC}},
  \href{https://doi.org/10.1103/PhysRevLett.115.081802}{\emph{Phys. Rev. Lett.}
  {\bfseries 115} (2015) 081802},
  [\href{https://arxiv.org/abs/1503.06834}{{\ttfamily 1503.06834}}].

\bibitem{Banerjee:2015gca}
S.~Banerjee, P.~S.~B. Dev, A.~Ibarra, T.~Mandal and M.~Mitra, \emph{{Prospects
  of Heavy Neutrino Searches at Future Lepton Colliders}},
  \href{https://doi.org/10.1103/PhysRevD.92.075002}{\emph{Phys. Rev.}
  {\bfseries D92} (2015) 075002},
  [\href{https://arxiv.org/abs/1503.05491}{{\ttfamily 1503.05491}}].

\bibitem{Arganda:2015ija}
E.~Arganda, M.~J. Herrero, X.~Marcano and C.~Weiland, \emph{{Exotic
  \ensuremath{\mu}\ensuremath{\tau}jj events from heavy ISS neutrinos at the
  LHC}}, \href{https://doi.org/10.1016/j.physletb.2015.11.013}{\emph{Phys.
  Lett. B} {\bfseries 752} (2016) 46--50},
  [\href{https://arxiv.org/abs/1508.05074}{{\ttfamily 1508.05074}}].

\bibitem{Das:2015toa}
A.~Das and N.~Okada, \emph{{Improved bounds on the heavy neutrino productions
  at the LHC}}, \href{https://doi.org/10.1103/PhysRevD.93.033003}{\emph{Phys.
  Rev.} {\bfseries D93} (2016) 033003},
  [\href{https://arxiv.org/abs/1510.04790}{{\ttfamily 1510.04790}}].

\bibitem{Gago:2015vma}
A.~M. Gago, P.~Hern\'andez, J.~Jones-P\'erez, M.~Losada and A.~Moreno
  Brice\~no, \emph{{Probing the Type I Seesaw Mechanism with Displaced Vertices
  at the LHC}},
  \href{https://doi.org/10.1140/epjc/s10052-015-3693-1}{\emph{Eur. Phys. J. C}
  {\bfseries 75} (2015) 470},
  [\href{https://arxiv.org/abs/1505.05880}{{\ttfamily 1505.05880}}].

\bibitem{Accomando:2016rpc}
E.~Accomando, L.~Delle~Rose, S.~Moretti, E.~Olaiya and C.~H.
  Shepherd-Themistocleous, \emph{{Novel SM-like Higgs decay into displaced
  heavy neutrino pairs in $U(1)'$ models}},
  \href{https://doi.org/10.1007/JHEP04(2017)081}{\emph{JHEP} {\bfseries 04}
  (2017) 081}, [\href{https://arxiv.org/abs/1612.05977}{{\ttfamily
  1612.05977}}].

\bibitem{Degrande:2016aje}
C.~Degrande, O.~Mattelaer, R.~Ruiz and J.~Turner, \emph{{Fully-Automated
  Precision Predictions for Heavy Neutrino Production Mechanisms at Hadron
  Colliders}}, \href{https://doi.org/10.1103/PhysRevD.94.053002}{\emph{Phys.
  Rev.} {\bfseries D94} (2016) 053002},
  [\href{https://arxiv.org/abs/1602.06957}{{\ttfamily 1602.06957}}].

\bibitem{Mitra:2016kov}
M.~Mitra, R.~Ruiz, D.~J. Scott and M.~Spannowsky, \emph{{Neutrino Jets from
  High-Mass $W_R$ Gauge Bosons in TeV-Scale Left-Right Symmetric Models}},
  \href{https://doi.org/10.1103/PhysRevD.94.095016}{\emph{Phys. Rev.}
  {\bfseries D94} (2016) 095016},
  [\href{https://arxiv.org/abs/1607.03504}{{\ttfamily 1607.03504}}].

\bibitem{Ruiz:2017yyf}
R.~Ruiz, M.~Spannowsky and P.~Waite, \emph{{Heavy neutrinos from gluon
  fusion}}, \href{https://doi.org/10.1103/PhysRevD.96.055042}{\emph{Phys. Rev.}
  {\bfseries D96} (2017) 055042},
  [\href{https://arxiv.org/abs/1706.02298}{{\ttfamily 1706.02298}}].

\bibitem{Dube:2017jgo}
S.~Dube, D.~Gadkari and A.~M. Thalapillil, \emph{{Lepton-Jets and Low-Mass
  Sterile Neutrinos at Hadron Colliders}},
  \href{https://doi.org/10.1103/PhysRevD.96.055031}{\emph{Phys. Rev.}
  {\bfseries D96} (2017) 055031},
  [\href{https://arxiv.org/abs/1707.00008}{{\ttfamily 1707.00008}}].

\bibitem{Caputo:2017pit}
A.~Caputo, P.~Hern{\'a}ndez, J.~L{\'o}pez-Pav{\'o}n and J.~Salvado, \emph{{The
  seesaw portal in testable models of neutrino masses}},
  \href{https://doi.org/10.1007/JHEP06(2017)112}{\emph{JHEP} {\bfseries 06}
  (2017) 112}, [\href{https://arxiv.org/abs/1704.08721}{{\ttfamily
  1704.08721}}].

\bibitem{Antusch:2017hhu}
S.~Antusch, E.~Cazzato and O.~Fischer, \emph{{Sterile neutrino searches via
  displaced vertices at LHCb}},
  \href{https://doi.org/10.1016/j.physletb.2017.09.057}{\emph{Phys. Lett.}
  {\bfseries B774} (2017) 114--118},
  [\href{https://arxiv.org/abs/1706.05990}{{\ttfamily 1706.05990}}].

\bibitem{Das:2017pvt}
A.~Das, \emph{{Pair production of heavy neutrinos in next-to-leading order QCD
  at the hadron colliders in the inverse seesaw framework}},
  \href{https://doi.org/10.1142/S0217751X21500123}{\emph{Int. J. Mod. Phys. A}
  {\bfseries 36} (2021) 2150012},
  [\href{https://arxiv.org/abs/1701.04946}{{\ttfamily 1701.04946}}].

\bibitem{Das:2017gke}
A.~Das, P.~Konar and A.~Thalapillil, \emph{{Jet substructure shedding light on
  heavy Majorana neutrinos at the LHC}},
  \href{https://doi.org/10.1007/JHEP02(2018)083}{\emph{JHEP} {\bfseries 02}
  (2018) 083}, [\href{https://arxiv.org/abs/1709.09712}{{\ttfamily
  1709.09712}}].

\bibitem{Deppisch:2018eth}
F.~F. Deppisch, W.~Liu and M.~Mitra, \emph{{Long-lived Heavy Neutrinos from
  Higgs Decays}}, \href{https://doi.org/10.1007/JHEP08(2018)181}{\emph{JHEP}
  {\bfseries 08} (2018) 181},
  [\href{https://arxiv.org/abs/1804.04075}{{\ttfamily 1804.04075}}].

\bibitem{Liu:2018wte}
J.~Liu, Z.~Liu and L.-T. Wang, \emph{{Enhancing Long-Lived Particles Searches
  at the LHC with Precision Timing Information}},
  \href{https://doi.org/10.1103/PhysRevLett.122.131801}{\emph{Phys. Rev. Lett.}
  {\bfseries 122} (2019) 131801},
  [\href{https://arxiv.org/abs/1805.05957}{{\ttfamily 1805.05957}}].

\bibitem{Cottin:2018kmq}
G.~Cottin, J.~C. Helo and M.~Hirsch, \emph{{Searches for light sterile
  neutrinos with multitrack displaced vertices}},
  \href{https://doi.org/10.1103/PhysRevD.97.055025}{\emph{Phys. Rev.}
  {\bfseries D97} (2018) 055025},
  [\href{https://arxiv.org/abs/1801.02734}{{\ttfamily 1801.02734}}].

\bibitem{Cottin:2018nms}
G.~Cottin, J.~C. Helo and M.~Hirsch, \emph{{Displaced vertices as probes of
  sterile neutrino mixing at the LHC}},
  \href{https://doi.org/10.1103/PhysRevD.98.035012}{\emph{Phys. Rev.}
  {\bfseries D98} (2018) 035012},
  [\href{https://arxiv.org/abs/1806.05191}{{\ttfamily 1806.05191}}].

\bibitem{Dib:2018iyr}
C.~O. Dib, C.~S. Kim, N.~A. Neill and X.-B. Yuan, \emph{{Search for sterile
  neutrinos decaying into pions at the LHC}},
  \href{https://doi.org/10.1103/PhysRevD.97.035022}{\emph{Phys. Rev.}
  {\bfseries D97} (2018) 035022},
  [\href{https://arxiv.org/abs/1801.03624}{{\ttfamily 1801.03624}}].

\bibitem{Nemevsek:2018bbt}
M.~Nemev{\v s}ek, F.~Nesti and G.~Popara, \emph{{Keung-Senjanovi{\'c} process
  at the LHC: From lepton number violation to displaced vertices to invisible
  decays}}, \href{https://doi.org/10.1103/PhysRevD.97.115018}{\emph{Phys. Rev.}
  {\bfseries D97} (2018) 115018},
  [\href{https://arxiv.org/abs/1801.05813}{{\ttfamily 1801.05813}}].

\bibitem{Abada:2018sfh}
A.~Abada, N.~Bernal, M.~Losada and X.~Marcano, \emph{{Inclusive Displaced
  Vertex Searches for Heavy Neutral Leptons at the LHC}},
  \href{https://doi.org/10.1007/JHEP01(2019)093}{\emph{JHEP} {\bfseries 01}
  (2019) 093}, [\href{https://arxiv.org/abs/1807.10024}{{\ttfamily
  1807.10024}}].

\bibitem{Pascoli:2018rsg}
S.~Pascoli, R.~Ruiz and C.~Weiland, \emph{{Safe Jet Vetoes}},
  \href{https://doi.org/10.1016/j.physletb.2018.08.060}{\emph{Phys. Lett.}
  {\bfseries B786} (2018) 106},
  [\href{https://arxiv.org/abs/1805.09335}{{\ttfamily 1805.09335}}].

\bibitem{Antusch:2018bgr}
S.~Antusch, E.~Cazzato, O.~Fischer, A.~Hammad and K.~Wang, \emph{{Lepton Flavor
  Violating Dilepton Dijet Signatures from Sterile Neutrinos at Proton
  Colliders}}, \href{https://doi.org/10.1007/JHEP10(2018)067}{\emph{JHEP}
  {\bfseries 10} (2018) 067},
  [\href{https://arxiv.org/abs/1805.11400}{{\ttfamily 1805.11400}}].

\bibitem{Pascoli:2018heg}
S.~Pascoli, R.~Ruiz and C.~Weiland, \emph{{Heavy neutrinos with dynamic jet
  vetoes: multilepton searches at $ \sqrt{s}=14$, 27, and 100 TeV}},
  \href{https://doi.org/10.1007/JHEP06(2019)049}{\emph{JHEP} {\bfseries 06}
  (2019) 049}, [\href{https://arxiv.org/abs/1812.08750}{{\ttfamily
  1812.08750}}].

\bibitem{Drewes:2019fou}
M.~Drewes and J.~Hajer, \emph{{Heavy Neutrinos in displaced vertex searches at
  the LHC and HL-LHC}},
  \href{https://doi.org/10.1007/JHEP02(2020)070}{\emph{JHEP} {\bfseries 02}
  (2020) 070}, [\href{https://arxiv.org/abs/1903.06100}{{\ttfamily
  1903.06100}}].

\bibitem{Liu:2019ayx}
J.~Liu, Z.~Liu, L.-T. Wang and X.-P. Wang, \emph{{Seeking for sterile neutrinos
  with displaced leptons at the LHC}},
  \href{https://doi.org/10.1007/JHEP07(2019)159}{\emph{JHEP} {\bfseries 07}
  (2019) 159}, [\href{https://arxiv.org/abs/1904.01020}{{\ttfamily
  1904.01020}}].

\bibitem{Jones-Perez:2019plk}
J.~Jones-P\'erez, J.~Masias and J.~D. Ruiz-\'Alvarez, \emph{{Search for
  Long-Lived Heavy Neutrinos at the LHC with a VBF Trigger}},
  \href{https://doi.org/10.1140/epjc/s10052-020-8188-z}{\emph{Eur. Phys. J. C}
  {\bfseries 80} (2020) 642},
  [\href{https://arxiv.org/abs/1912.08206}{{\ttfamily 1912.08206}}].

\bibitem{Fuks:2020att}
B.~Fuks, J.~Neundorf, K.~Peters, R.~Ruiz and M.~Saimpert, \emph{{Majorana
  neutrinos in same-sign $W^\pm W^\pm$ scattering at the LHC: Breaking the TeV
  barrier}}, \href{https://doi.org/10.1103/PhysRevD.103.055005}{\emph{Phys.
  Rev. D} {\bfseries 103} (2021) 055005},
  [\href{https://arxiv.org/abs/2011.02547}{{\ttfamily 2011.02547}}].

\bibitem{Tastet:2021vwp}
J.-L. Tastet, O.~Ruchayskiy and I.~Timiryasov, \emph{{Reinterpreting the ATLAS
  bounds on heavy neutral leptons in a realistic neutrino oscillation model}},
  \href{https://doi.org/10.1007/JHEP12(2021)182}{\emph{JHEP} {\bfseries 12}
  (2021) 182}, [\href{https://arxiv.org/abs/2107.12980}{{\ttfamily
  2107.12980}}].

\bibitem{LHCReinterpretationForum:2020xtr}
{\scshape LHC Reinterpretation Forum} collaboration, W.~Abdallah et~al.,
  \emph{{Reinterpretation of LHC Results for New Physics: Status and
  Recommendations after Run 2}},
  \href{https://doi.org/10.21468/SciPostPhys.9.2.022}{\emph{SciPost Phys.}
  {\bfseries 9} (2020) 022},
  [\href{https://arxiv.org/abs/2003.07868}{{\ttfamily 2003.07868}}].

\bibitem{Ibarra:2003up}
A.~Ibarra and G.~G. Ross, \emph{{Neutrino phenomenology: The Case of two
  right-handed neutrinos}},
  \href{https://doi.org/10.1016/j.physletb.2004.04.037}{\emph{Phys. Lett. B}
  {\bfseries 591} (2004) 285--296},
  [\href{https://arxiv.org/abs/hep-ph/0312138}{{\ttfamily hep-ph/0312138}}].

\bibitem{Gavela:2009cd}
M.~B. Gavela, T.~Hambye, D.~Hernandez and P.~Hernandez, \emph{{Minimal Flavour
  Seesaw Models}},
  \href{https://doi.org/10.1088/1126-6708/2009/09/038}{\emph{JHEP} {\bfseries
  09} (2009) 038}, [\href{https://arxiv.org/abs/0906.1461}{{\ttfamily
  0906.1461}}].

\bibitem{Drewes:2022akb}
M.~Drewes, J.~Klari\'c and J.~L\'opez-Pav\'on, \emph{{New Benchmark Models for
  Heavy Neutral Lepton Searches}},
  \href{https://arxiv.org/abs/2207.02742}{{\ttfamily 2207.02742}}.

\bibitem{Das:2017hmg}
A.~Das, P.~S.~B. Dev and R.~N. Mohapatra, \emph{{Same Sign versus Opposite Sign
  Dileptons as a Probe of Low Scale Seesaw Mechanisms}},
  \href{https://doi.org/10.1103/PhysRevD.97.015018}{\emph{Phys. Rev. D}
  {\bfseries 97} (2018) 015018},
  [\href{https://arxiv.org/abs/1709.06553}{{\ttfamily 1709.06553}}].

\bibitem{Abada:2019bac}
A.~Abada, C.~Hati, X.~Marcano and A.~M. Teixeira, \emph{{Interference effects
  in LNV and LFV semileptonic decays: the Majorana hypothesis}},
  \href{https://doi.org/10.1007/JHEP09(2019)017}{\emph{JHEP} {\bfseries 09}
  (2019) 017}, [\href{https://arxiv.org/abs/1904.05367}{{\ttfamily
  1904.05367}}].

\bibitem{Najafi:2020dkp}
F.~Najafi, J.~Kumar and D.~London, \emph{{CP violation in rare
  lepton-number-violating $W$ decays at the LHC}},
  \href{https://doi.org/10.1007/JHEP04(2021)021}{\emph{JHEP} {\bfseries 04}
  (2021) 021}, [\href{https://arxiv.org/abs/2011.03686}{{\ttfamily
  2011.03686}}].

\bibitem{Cai:2017mow}
Y.~Cai, T.~Han, T.~Li and R.~Ruiz, \emph{{Lepton Number Violation: Seesaw
  Models and Their Collider Tests}},
  \href{https://doi.org/10.3389/fphy.2018.00040}{\emph{Front. in Phys.}
  {\bfseries 6} (2018) 40}, [\href{https://arxiv.org/abs/1711.02180}{{\ttfamily
  1711.02180}}].

\bibitem{Batell:2022ubw}
{\scshape \$nu\$-Test} collaboration, B.~Batell, T.~Ghosh and K.~Xie,
  \emph{{Heavy Neutral Lepton Searches at the Electron-Ion Collider: A Snowmass
  Whitepaper}},  in \emph{{2022 Snowmass Summer Study}}, 3, 2022,
  \href{https://arxiv.org/abs/2203.06705}{{\ttfamily 2203.06705}}.

\bibitem{Abdullahi:2022jlv}
A.~M. Abdullahi et~al., \emph{{The Present and Future Status of Heavy Neutral
  Leptons}},  in \emph{{2022 Snowmass Summer Study}}, 3, 2022,
  \href{https://arxiv.org/abs/2203.08039}{{\ttfamily 2203.08039}}.

\bibitem{CMS:2018jxx}
{\scshape CMS} collaboration, A.~M. Sirunyan et~al., \emph{{Search for heavy
  Majorana neutrinos in same-sign dilepton channels in proton-proton collisions
  at $ \sqrt{s}=13 $ TeV}},
  \href{https://doi.org/10.1007/JHEP01(2019)122}{\emph{JHEP} {\bfseries 01}
  (2019) 122}, [\href{https://arxiv.org/abs/1806.10905}{{\ttfamily
  1806.10905}}].

\bibitem{Antusch:2016ejd}
S.~Antusch, E.~Cazzato and O.~Fischer, \emph{{Sterile neutrino searches at
  future $e^-e^+$, $pp$, and $e^-p$ colliders}},
  \href{https://doi.org/10.1142/S0217751X17500786}{\emph{Int. J. Mod. Phys. A}
  {\bfseries 32} (2017) 1750078},
  [\href{https://arxiv.org/abs/1612.02728}{{\ttfamily 1612.02728}}].

\bibitem{ATLAS:2011izm}
{\scshape ATLAS} collaboration, G.~Aad et~al., \emph{{Inclusive search for
  same-sign dilepton signatures in $pp$ collisions at $\sqrt{s}=7$ TeV with the
  ATLAS detector}}, \href{https://doi.org/10.1007/JHEP10(2011)107}{\emph{JHEP}
  {\bfseries 10} (2011) 107},
  [\href{https://arxiv.org/abs/1108.0366}{{\ttfamily 1108.0366}}].

\bibitem{ATLAS:2012ak}
{\scshape ATLAS} collaboration, G.~Aad et~al., \emph{{Search for heavy
  neutrinos and right-handed $W$ bosons in events with two leptons and jets in
  $pp$ collisions at $\sqrt{s}=7$ TeV with the ATLAS detector}},
  \href{https://doi.org/10.1140/epjc/s10052-012-2056-4}{\emph{Eur. Phys. J. C}
  {\bfseries 72} (2012) 2056},
  [\href{https://arxiv.org/abs/1203.5420}{{\ttfamily 1203.5420}}].

\bibitem{CMS:2012zv}
{\scshape CMS} collaboration, S.~Chatrchyan et~al., \emph{{Search for Heavy
  Neutrinos and $W_R$ Bosons with Right-Handed Couplings in a Left-Right
  Symmetric Model in pp Collisions at $\sqrt{s}$ = 7 TeV}},
  \href{https://doi.org/10.1103/PhysRevLett.109.261802}{\emph{Phys. Rev. Lett.}
  {\bfseries 109} (2012) 261802},
  [\href{https://arxiv.org/abs/1210.2402}{{\ttfamily 1210.2402}}].

\bibitem{CMS:2014nrz}
{\scshape CMS} collaboration, V.~Khachatryan et~al., \emph{{Search for Heavy
  Neutrinos and $\mathrm {W}$ Bosons with Right-Handed Couplings in
  Proton-Proton Collisions at $\sqrt{s} = 8\,\text {TeV} $}},
  \href{https://doi.org/10.1140/epjc/s10052-014-3149-z}{\emph{Eur. Phys. J. C}
  {\bfseries 74} (2014) 3149},
  [\href{https://arxiv.org/abs/1407.3683}{{\ttfamily 1407.3683}}].

\bibitem{CMS:2018agk}
{\scshape CMS} collaboration, A.~M. Sirunyan et~al., \emph{{Search for a heavy
  right-handed W boson and a heavy neutrino in events with two same-flavor
  leptons and two jets at $\sqrt{s}=$ 13 TeV}},
  \href{https://doi.org/10.1007/JHEP05(2018)148}{\emph{JHEP} {\bfseries 05}
  (2018) 148}, [\href{https://arxiv.org/abs/1803.11116}{{\ttfamily
  1803.11116}}].

\bibitem{ATLAS:2018dcj}
{\scshape ATLAS} collaboration, M.~Aaboud et~al., \emph{{Search for heavy
  Majorana or Dirac neutrinos and right-handed $W$ gauge bosons in final states
  with two charged leptons and two jets at $ \sqrt{s}=13 $ TeV with the ATLAS
  detector}}, \href{https://doi.org/10.1007/JHEP01(2019)016}{\emph{JHEP}
  {\bfseries 01} (2019) 016},
  [\href{https://arxiv.org/abs/1809.11105}{{\ttfamily 1809.11105}}].

\bibitem{CMS:2018iye}
{\scshape CMS} collaboration, A.~M. Sirunyan et~al., \emph{{Search for heavy
  neutrinos and third-generation leptoquarks in hadronic states of two $\tau$
  leptons and two jets in proton-proton collisions at $\sqrt{s} =$ 13 TeV}},
  \href{https://doi.org/10.1007/JHEP03(2019)170}{\emph{JHEP} {\bfseries 03}
  (2019) 170}, [\href{https://arxiv.org/abs/1811.00806}{{\ttfamily
  1811.00806}}].

\bibitem{ATLAS:2019isd}
{\scshape ATLAS} collaboration, M.~Aaboud et~al., \emph{{Search for a
  right-handed gauge boson decaying into a high-momentum heavy neutrino and a
  charged lepton in $pp$ collisions with the ATLAS detector at $\sqrt{s}=13$
  TeV}}, \href{https://doi.org/10.1016/j.physletb.2019.134942}{\emph{Phys.
  Lett. B} {\bfseries 798} (2019) 134942},
  [\href{https://arxiv.org/abs/1904.12679}{{\ttfamily 1904.12679}}].

\bibitem{Moffat:2017feq}
K.~Moffat, S.~Pascoli and C.~Weiland, \emph{{Equivalence between massless
  neutrinos and lepton number conservation in fermionic singlet extensions of
  the Standard Model}},  \href{https://arxiv.org/abs/1712.07611}{{\ttfamily
  1712.07611}}.

\bibitem{Antusch:2017ebe}
S.~Antusch, E.~Cazzato and O.~Fischer, \emph{{Resolvable heavy
  neutrino\textendash{}antineutrino oscillations at colliders}},
  \href{https://doi.org/10.1142/S0217732319500615}{\emph{Mod. Phys. Lett. A}
  {\bfseries 34} (2019) 1950061},
  [\href{https://arxiv.org/abs/1709.03797}{{\ttfamily 1709.03797}}].

\bibitem{Drewes:2019byd}
M.~Drewes, J.~Klari\'c and P.~Klose, \emph{{On lepton number violation in heavy
  neutrino decays at colliders}},
  \href{https://doi.org/10.1007/JHEP11(2019)032}{\emph{JHEP} {\bfseries 11}
  (2019) 032}, [\href{https://arxiv.org/abs/1907.13034}{{\ttfamily
  1907.13034}}].

\bibitem{Fernandez-Martinez:2022gsu}
E.~Fern\'andez-Mart\'\i{}nez, X.~Marcano and D.~Naredo-Tuero, \emph{{HNL mass
  degeneracy: implications for low-scale seesaws, LNV at colliders and
  leptogenesis}},  \href{https://arxiv.org/abs/2209.04461}{{\ttfamily
  2209.04461}}.

\bibitem{LHCb:2020wxx}
{\scshape LHCb} collaboration, R.~Aaij et~al., \emph{{Search for heavy neutral
  leptons in $W^+\to\mu^{+}\mu^{\pm}\text{jet}$ decays}},
  \href{https://doi.org/10.1140/epjc/s10052-021-08973-5}{\emph{Eur. Phys. J. C}
  {\bfseries 81} (2021) 248},
  [\href{https://arxiv.org/abs/2011.05263}{{\ttfamily 2011.05263}}].

\bibitem{ATLAS:2019kpx}
{\scshape ATLAS} collaboration, G.~Aad et~al., \emph{{Search for heavy neutral
  leptons in decays of $W$ bosons produced in 13 TeV $pp$ collisions using
  prompt and displaced signatures with the ATLAS detector}},
  \href{https://doi.org/10.1007/JHEP10(2019)265}{\emph{JHEP} {\bfseries 10}
  (2019) 265}, [\href{https://arxiv.org/abs/1905.09787}{{\ttfamily
  1905.09787}}].

\bibitem{ATLAS:2022atq}
{\scshape ATLAS} collaboration, \emph{{Search for heavy neutral leptons in
  decays of $W$ bosons using a dilepton displaced vertex in $\sqrt{s}=13$ TeV
  $pp$ collisions with the ATLAS detector}},
  \href{https://arxiv.org/abs/2204.11988}{{\ttfamily 2204.11988}}.

\bibitem{CMS:2022fut}
{\scshape CMS} collaboration, A.~Tumasyan et~al., \emph{{Search for long-lived
  heavy neutral leptons with displaced vertices in proton-proton collisions at
  $ \sqrt{\mathrm{s}} $ =13 TeV}},
  \href{https://doi.org/10.1007/JHEP07(2022)081}{\emph{JHEP} {\bfseries 07}
  (2022) 081}, [\href{https://arxiv.org/abs/2201.05578}{{\ttfamily
  2201.05578}}].

\bibitem{CMS:2012wqj}
{\scshape CMS} collaboration, S.~Chatrchyan et~al., \emph{{Search for heavy
  Majorana Neutrinos in $\mu^{\pm}\mu^{\pm} +$ Jets and $e^{\pm}e^{\pm} +$ Jets
  Events in pp Collisions at $\sqrt{s} =$ 7 TeV}},
  \href{https://doi.org/10.1016/j.physletb.2012.09.012}{\emph{Phys. Lett. B}
  {\bfseries 717} (2012) 109--128},
  [\href{https://arxiv.org/abs/1207.6079}{{\ttfamily 1207.6079}}].

\bibitem{CMS:2015qur}
{\scshape CMS} collaboration, V.~Khachatryan et~al., \emph{{Search for heavy
  Majorana neutrinos in $\mu^\pm \mu^\pm+$ jets events in proton-proton
  collisions at $\sqrt{s}$ = 8 TeV}},
  \href{https://doi.org/10.1016/j.physletb.2015.06.070}{\emph{Phys. Lett. B}
  {\bfseries 748} (2015) 144--166},
  [\href{https://arxiv.org/abs/1501.05566}{{\ttfamily 1501.05566}}].

\bibitem{CMS:2016aro}
{\scshape CMS} collaboration, V.~Khachatryan et~al., \emph{{Search for heavy
  Majorana neutrinos in e$^\pm$e$^\pm$+ jets and e$^\pm$ $\mu^\pm$+ jets events
  in proton-proton collisions at $ \sqrt{s}=8 $ TeV}},
  \href{https://doi.org/10.1007/JHEP04(2016)169}{\emph{JHEP} {\bfseries 04}
  (2016) 169}, [\href{https://arxiv.org/abs/1603.02248}{{\ttfamily
  1603.02248}}].

\bibitem{ATLAS:2015gtp}
{\scshape ATLAS} collaboration, G.~Aad et~al., \emph{{Search for heavy Majorana
  neutrinos with the ATLAS detector in pp collisions at $ \sqrt{s}=8 $ TeV}},
  \href{https://doi.org/10.1007/JHEP07(2015)162}{\emph{JHEP} {\bfseries 07}
  (2015) 162}, [\href{https://arxiv.org/abs/1506.06020}{{\ttfamily
  1506.06020}}].

\bibitem{CMS:2018iaf}
{\scshape CMS} collaboration, A.~M. Sirunyan et~al., \emph{{Search for heavy
  neutral leptons in events with three charged leptons in proton-proton
  collisions at $\sqrt{s} =$ 13 TeV}},
  \href{https://doi.org/10.1103/PhysRevLett.120.221801}{\emph{Phys. Rev. Lett.}
  {\bfseries 120} (2018) 221801},
  [\href{https://arxiv.org/abs/1802.02965}{{\ttfamily 1802.02965}}].

\bibitem{Fernandez-Martinez:2016lgt}
E.~Fernandez-Martinez, J.~Hernandez-Garcia and J.~Lopez-Pavon, \emph{{Global
  constraints on heavy neutrino mixing}},
  \href{https://doi.org/10.1007/JHEP08(2016)033}{\emph{JHEP} {\bfseries 08}
  (2016) 033}, [\href{https://arxiv.org/abs/1605.08774}{{\ttfamily
  1605.08774}}].

\bibitem{Blondel:2014bra}
{\scshape FCC-ee study Team} collaboration, A.~Blondel, E.~Graverini, N.~Serra
  and M.~Shaposhnikov, \emph{{Search for Heavy Right Handed Neutrinos at the
  FCC-ee}},
  \href{https://doi.org/10.1016/j.nuclphysbps.2015.09.304}{\emph{Nucl. Part.
  Phys. Proc.} {\bfseries 273-275} (2016) 1883--1890},
  [\href{https://arxiv.org/abs/1411.5230}{{\ttfamily 1411.5230}}].

\bibitem{DELPHI:1996qcc}
{\scshape DELPHI} collaboration, P.~Abreu et~al., \emph{{Search for neutral
  heavy leptons produced in Z decays}},
  \href{https://doi.org/10.1007/s002880050370}{\emph{Z. Phys. C} {\bfseries 74}
  (1997) 57--71}.

\bibitem{L3:2001zfe}
{\scshape L3} collaboration, P.~Achard et~al., \emph{{Search for heavy
  isosinglet neutrino in $e^{+} e^{-}$ annihilation at LEP}},
  \href{https://doi.org/10.1016/S0370-2693(01)00993-5}{\emph{Phys. Lett. B}
  {\bfseries 517} (2001) 67--74},
  [\href{https://arxiv.org/abs/hep-ex/0107014}{{\ttfamily hep-ex/0107014}}].

\bibitem{Bondarenko:2018ptm}
K.~Bondarenko, A.~Boyarsky, D.~Gorbunov and O.~Ruchayskiy, \emph{{Phenomenology
  of GeV-scale Heavy Neutral Leptons}},
  \href{https://doi.org/10.1007/JHEP11(2018)032}{\emph{JHEP} {\bfseries 11}
  (2018) 032}, [\href{https://arxiv.org/abs/1805.08567}{{\ttfamily
  1805.08567}}].

\bibitem{Ibarra:2010xw}
A.~Ibarra, E.~Molinaro and S.~T. Petcov, \emph{{TeV Scale See-Saw Mechanisms of
  Neutrino Mass Generation, the Majorana Nature of the Heavy Singlet Neutrinos
  and $(\beta\beta)_{0\nu}$-Decay}},
  \href{https://doi.org/10.1007/JHEP09(2010)108}{\emph{JHEP} {\bfseries 09}
  (2010) 108}, [\href{https://arxiv.org/abs/1007.2378}{{\ttfamily 1007.2378}}].

\bibitem{Abada:2014vea}
A.~Abada and M.~Lucente, \emph{{Looking for the minimal inverse seesaw
  realisation}},
  \href{https://doi.org/10.1016/j.nuclphysb.2014.06.003}{\emph{Nucl. Phys. B}
  {\bfseries 885} (2014) 651--678},
  [\href{https://arxiv.org/abs/1401.1507}{{\ttfamily 1401.1507}}].

\bibitem{Pilaftsis:2003gt}
A.~Pilaftsis and T.~E.~J. Underwood, \emph{{Resonant leptogenesis}},
  \href{https://doi.org/10.1016/j.nuclphysb.2004.05.029}{\emph{Nucl. Phys. B}
  {\bfseries 692} (2004) 303--345},
  [\href{https://arxiv.org/abs/hep-ph/0309342}{{\ttfamily hep-ph/0309342}}].

\bibitem{Akhmedov:1998qx}
E.~K. Akhmedov, V.~A. Rubakov and A.~Y. Smirnov, \emph{{Baryogenesis via
  neutrino oscillations}},
  \href{https://doi.org/10.1103/PhysRevLett.81.1359}{\emph{Phys. Rev. Lett.}
  {\bfseries 81} (1998) 1359--1362},
  [\href{https://arxiv.org/abs/hep-ph/9803255}{{\ttfamily hep-ph/9803255}}].

\bibitem{Brass:2019hvu}
S.~Bra\ss{}, W.~Kilian, T.~Ohl, J.~Reuter, V.~Rothe and P.~Stienemeier,
  \emph{{Precision Monte Carlo simulations with WHIZARD}},
  \href{https://doi.org/10.23731/CYRM-2020-003.205}{\emph{CERN Yellow Reports:
  Monographs} {\bfseries 3} (2020) 205--210}.

\bibitem{Shtabovenko:2016sxi}
V.~Shtabovenko, R.~Mertig and F.~Orellana, \emph{{New Developments in FeynCalc
  9.0}}, \href{https://doi.org/10.1016/j.cpc.2016.06.008}{\emph{Comput. Phys.
  Commun.} {\bfseries 207} (2016) 432--444},
  [\href{https://arxiv.org/abs/1601.01167}{{\ttfamily 1601.01167}}].

\end{thebibliography}\endgroup

\end{document}